# Advances in perovskite nanocrystals and nanocomposites for scintillation applications

*Abhinav Anand[1,2], Matteo L. Zaffalon[1], Andrea Erroi[1], Francesca Cova[1], Francesco Carulli[1]* and Sergio Brovelli[1]**

[1]*Dipartimento di Scienza dei Materiali, Università degli Studi di Milano-Bicocca, via R. Cozzi 55, 20125, Milano, Italy.*
[2]*Department of Physics, School of Advanced Sciences, Vellore Institute of Technology, Vellore, Tamil Nadu, 632007, India.*

*\*francesco.carulli@unimib.it sergio.brovelli@unimib.it*

**Abstract**

In recent years, the field of radiation detection has witnessed a paradigm shift with the emergence of plastic scintillators incorporating perovskite nanocrystals (PNCs). This innovative class of scintillators not only capitalizes on the superior luminescent properties of PNCs but also harnesses the flexibility and processability of polymers. This review explores the intricate landscape of synthesizing and fabricating scintillating PNCs and nanocomposites, delving into the methods employed in their production. From solution-based methods to innovative solid-state approaches, the synthesis of PNCs for scintillators application is explored comprehensively. Furthermore, embedding strategies within polymeric matrices are scrutinized, shedding light on the various techniques utilized to achieve optimal dispersion and compatibility. The evaluation of the final nanocomposites is finally discussed, with a particular emphasis on their scintillating performance and radiation hardness. Through a meticulous exploration of synthesis methodologies, embedding techniques, and performance assessments, this review aims to provide a multilayered understanding of the state-of-the-art in PNCs-based nanoscintillators.

The detection of high-energy photons (X or γ) and particles (α, β or neutrons), commonly referred to as ionizing radiation, is at the heart of many strategic applications in modern science and technology, including medical diagnostics[1, 2] and therapy[3], non-destructive industrial inspection[4], border and national security[5, 6], deep space exploration [7] and high-energy physics research[8]. The key principle in all these applications is the conversion of all or part of the energy released by incident ionizing radiation into a target material, which can occur by various mechanisms such as Coulomb collisions, Compton scattering, photoelectric effect and pair production, depending on the type and energy of the radiation (as schematically shown in **Figures 1a**). These phenomena typically lead to the promotion of electrons from the inner states of atoms interacting with ionizing radiation and the subsequent formation of a large number of secondary electrons (and holes) that thermalize to the bottom (top) of the conduction (valence) band of the collided material. There are two main architectures for radiation detection (scheme in **Figure 1b**): *direct detectors*, in which the charges released by the ionizing radiation are read as an electrical current, and *indirect detectors*, which rely on the optical readout by photomultiplier tubes (PMTs) or photodiodes (typically Si-based) of UV or visible photons produced by the radiative recombination of the generated charges in a process known as 'scintillation'. There are several essential features that determine the efficiency of a scintillator detector: (*i*) the probability of interaction with ionizing radiation, which depends on the effective atomic number (as $\sigma \propto Z^n$, with $1<n<5$, depending on the type of interaction[9]) and density of a scintillating material and ultimately determines its stopping power, σ; (*ii*) the light yield (LY), defined as the number of scintillation photons emitted per unit of deposited energy[9] (for example, interaction with 1 MeV gamma ray typically produces ~$10^3$-$10^5$ photons per event[10]), which quantifies the ability to convert the generated charges into photons, (*iii*) the scintillation lifetime, which is critical for applications that rely on the timing of events, such as time-of-flight (TOF) based technologies, as it allows operation at high rates and discrimination of scintillation events without detrimental pile-up effects; (*iv*) radiation hardness, which expresses the scintillator's resistance to prolonged exposure to ionizing radiation. Although scintillators are designed to measure ionizing radiation, radiation damage can occur when exposed to high levels of radiation, resulting in degradation of light production and collection due to the formation of non-radiative traps and defects as well as mechanical damage. Radiation hardness is not a strictly defined quantity, but it is often used to assess the maximum dose tolerated by actual detectors, which relates to the ability of a material to maintain its characteristics unchanged after irradiation. Currently, most scintillating devices are based on inorganic crystals, which are efficient and relatively fast, but are grown using expensive and energy-intensive techniques (molten salt preparation) and are limited in manufacturing scalability[11].

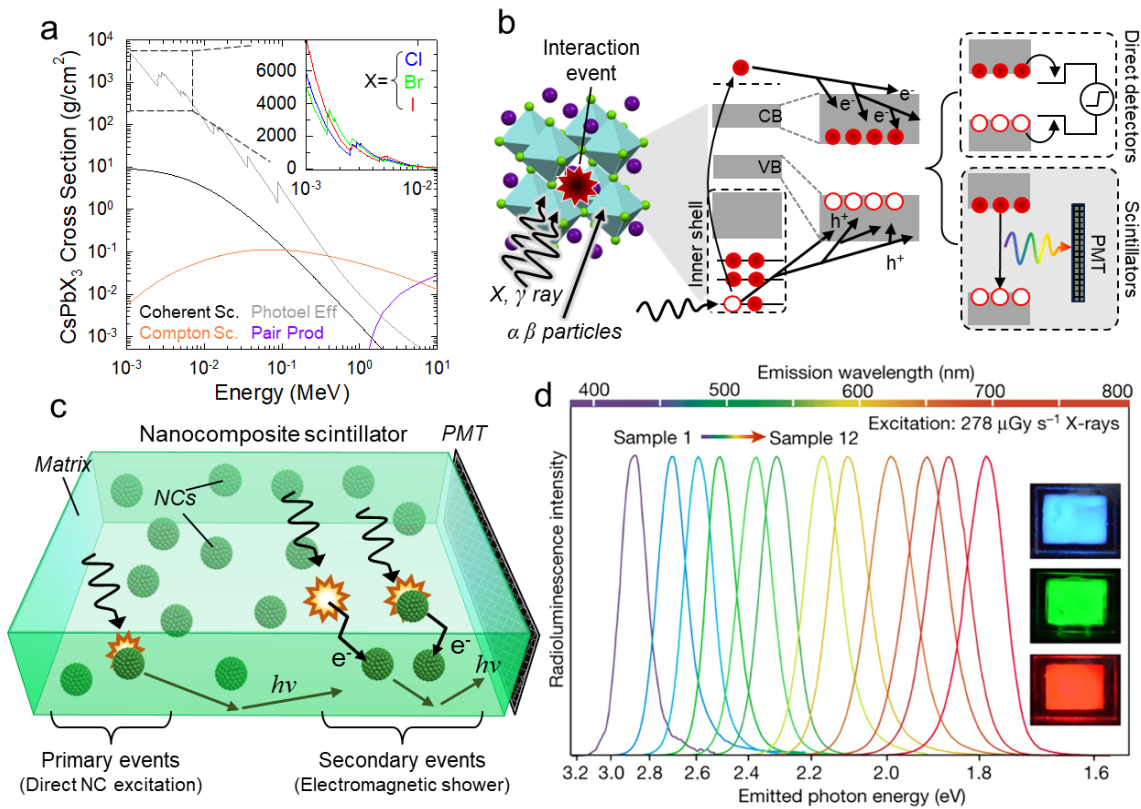

**Figure 1:** a) Different types of interaction mechanisms of ionizing radiation with matter. Inset: Photon cross section of $CsPbX_3$ PNCs (X=Cl, Br or I) as a function of ionizing photon energy. b) Representation of the process of interaction with ionizing radiation, generated in an all-inorganic perovskite lattice with a cubic crystal structure and functioning of direct detectors and scintillators. c) Structure of a nanoscintillator and physical processes involved in the scintillation mechanism. d) Tuneable RL of PNCs under X-ray excitation (dose rate 278 μGy s$^{-1}$ at 50 kV). The material compositions of samples are CsPbX (with X=$Cl_3$, $Cl_2Br$, $Cl_{1.5}Br_{1.5}$, $ClBr_2$, $Cl_{2.5}Br_{0.5}$, $Br_3$, $Br_2I$, $Br_{1.8}I_{1.2}$, $Br_{1.5}I_{1.5}$, $Br_{1.2}I_{1.8}$, $BrI_2$ and $I_3$, from 1 to 12 respectively). The insets show photographs of the thin-film samples 3, 6 and 9. Reproduced or adapted with permission from Ref ([12]). Copyright 2018, Springer.

On the other hand, plastic scintillators, consisting of polymeric matrixes doped with organic dyes, are customizable in size and shape, easily scalable at low cost and lightweight, but their interaction with ionizing radiation is intrinsically limited by their low density and they suffer from poor radiation stability[13, 14].

In recent years, nanocomposite scintillators have emerged as a novel approach to bridge the gap between conventional inorganic single crystals and plastic scintillators. This novel design, consisting of plastic matrixes doped or coated with scintillating nanocrystals (NCs) or nanoparticles, takes advantage of the heavy element composition and radiation hardness of inorganic crystals combined with the flexibility and low fabrication cost of plastic scintillators[15-17] (**Figure 1c**). Within this framework, the last decade has seen significant progress in the development of nanocomposite materials, introducing a wide range of high-Z nanoparticles such as $BaF_2$[18], $HfO_2$[19], rare earth fluorides and oxides[20, 21], as well as colloidal semiconductor NCs of various compositions and shapes[22-27]. Within the latter class of materials, lead halide perovskite NCs (PNCs) of the general formula $APbX_3$ (where A is a monovalent anion and X is a halogen) are of particular interest for scintillation applications[12, 28-31] due to their unique combination of scalable fabrication, heavy-element composition and advantageous photophysics, including size and halide composition - emission wavelength control[12, 32] to match the peak efficiency of photodetectors (**Figure 1d**) and unparalleled tolerance to point defects, resulting in high emission efficiency without complex post-processing or overcoating. PNCs can be synthesized using low-temperature, high-yield solution methods and are easy to integrate on an industrial scale, characteristics that make them ideal candidates for integration into nanocomposite scintillator manufacturing processes. In recent years, research efforts have greatly increased our knowledge of nanoscintillators by expanding the composition of PNCs used, developing new nanocomposite fabrication approaches, and correlating the physical properties of PNCs with scintillation performance. This review explores recent advances in scintillating PNCs, with a particular focus on their integration into polymeric nanocomposites for enhanced plastic scintillators. The different types of PNCs, polymer embedding strategies, scintillation properties and performance of these nanocomposites are examined, providing insights into their potential to revolutionize ionizing radiation detection technologies.

- **PNCs as nanoscintillators**

Lead halide perovskites for ionizing radiation detection have already been widely used in the form of thin films[19, 33-36] or single crystals[11, 37-40] in direct-architecture devices[41], which aimed to exploit both their high interaction with incident radiation due to their heavy-element-based composition and their excellent charge-carrier mobility[42, 43]. These remarkable results showed that lead halide perovskites are promising materials for detecting ionizing radiation and provided fresh motivation for further research. Colloidal synthesis strategies for PNCs developed in the mid-2010s have made it possible to produce these materials in the form of NCs using scalable processes and with a high degree of control over their optical properties, that, together with their already known good interaction with high-energy photons and resilience to ionizing radiation, has broadened the applicability of halide perovskites in the field of ionizing radiation detection also as scintillators.

To date, the most popular methods for the synthesis of PNCs featuring excellent optical features[44, 45] are ligand-assisted reprecipitation (LARP) and hot injection (HI). LARP consists of ultrafast crystallization of solubilized precursor salts in polar solvents (controlled by appropriate dosing of ligands) triggered by their injection into a non-polar solvent. By virtue of its scalable nature, the LARP method is particularly suitable for synthesizing PNCs in large volumes[46-48]. HI, on the other hand, involves the hot addition of one of the perovskite components (typically the monovalent cation) into a solution containing the other precursors. High-quality $CsPbBr_3$ PNCs with near-unity PLQY are normally synthesized by the HI approach[32, 48, 49]. However, compared to hot injection, the LARP approach can produce multigram quantities in a single batch, which is beneficial for fabricating large scintillators[12, 28]. This room temperature procedure however, suffers from concentration gradients resulting in low optical performance[50, 51] and often requires post synthetic treatment to yield PLQY >80%[52-55]. A remarkable step forward towards high control of PNCs features during their synthesis was recently reported by Akkerman et al.[51] who showed that highly monodispersed spheroidal lead halide PNCs can be obtained by controlling conversion of precursors into reactive species: this leads to the separation of the nucleation from the growth phases and enables to finely control the PNCs dimensions with the reaction time.

One of the first reports on the use of all-inorganic PNCs as scintillators was published in 2017 by Chen et al.[56]. Here, they showed that hot-injected $CsPbBr_3$ PNCs can directly interact with X-rays to produce radioluminescence effects. The RL properties of $CsPbBr_3$ PNCs were characterized under different solvents, concentrations, and X-ray irradiation environments (**Figure 2a**). Shortly thereafter, Heo et al.[28] in 2018 reported one of the first direct comparisons between all-inorganic PNCs and commercial bulk scintillators. They successfully demonstrated $CsPbBr_3$ PNCs-based scintillator X-ray detectors with high long-term stability (preserving 96% of initial efficiency under 40 Gy·s$^{-1}$ exposure), fast response time ($\tau_{avg}^{PL}$ =2.87 ns) and high resolution (9.8 lp mm$^{-1}$). Remarkably, the $CsPbBr_3$ PNCs scintillator exhibited a higher light output/absorptivity ratio than conventional gadolinium oxysulfide (GOS) crystals (45% vs 12%, expressed as light output power density/X-ray absorption), which are typically used in X-ray imaging screens. Equally important, as a result of the direct band-edge excitonic structure, the time response of RL was five times faster than that of GOS (**Figure 2b**) ($\tau_{avg}^{PL}$ =15.99 ns). This, combined with the fabrication of films with reduced size dispersion and suppressed scattering, allowed higher spatial resolutions to be achieved, as shown in **Figure 2c**. In addition, the $CsPbBr_3$ PNCs scintillators were reported to have comparable stability to the conventional GOS scintillator at an irradiation rate of 40 Gy·s$^{-1}$.

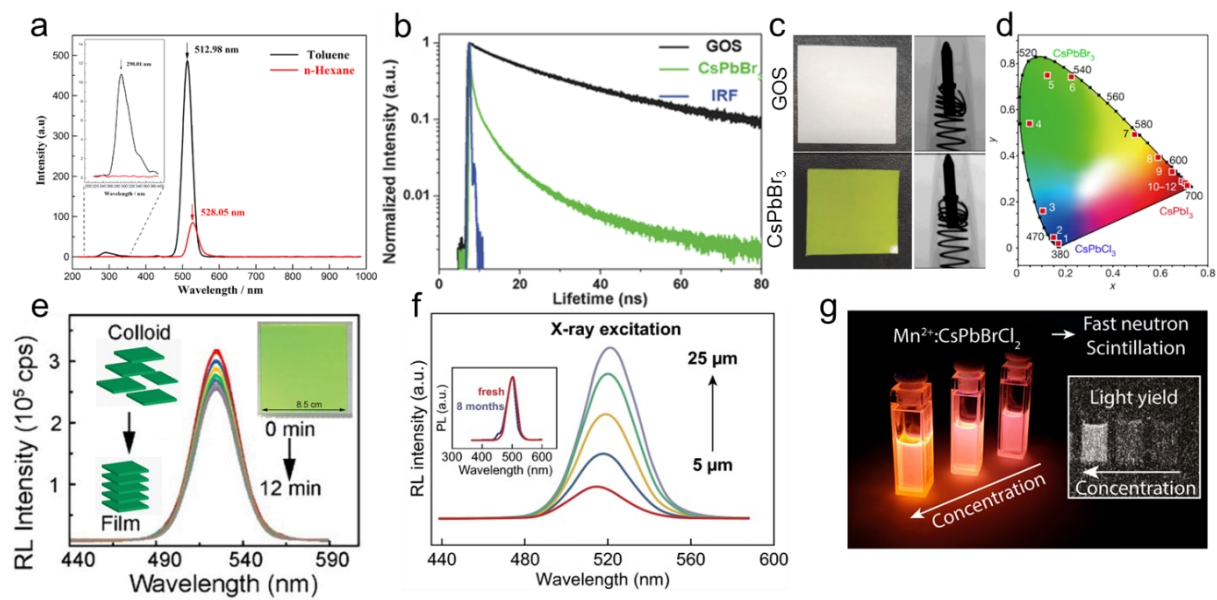

**Figure 2:** a) RL spectra of a toluene and n-hexane solution of $CsPbBr_3$ PNCs (10 mg/mL) under irradiation with 60 kV and 1 mA. Reproduced or adapted with permission from ref. ([56]). Copyright 2017, Springer. b) PL dynamics of $CsPbBr_3$ PNCs and conventional GOS scintillators-based X-ray photodetector excited with 285 nm UV light. c) Photographs of conventional GOS and $CsPbBr_3$ PNCs scintillator films under room light and corresponding X-ray images of ball-points pens taken by conventional GOS- and $CsPbBr_3$ PNCs- scintillator based X-ray detectors. Reproduced or adapted with permission from Ref ([28]). Copyright 2018, Wiley. d) CIE (Commission Internationale de l'Eclairage) chromaticity coordinates of the X-ray-induced visible emissions measured for metal halide PNCs samples with different composition. Reproduced or adapted with permission from Ref ([12]). Copyright 2018, Springer. e) RL spectra monitored during the transformation from a colloid to a solid in 12 min. Inset: Schematic showing the self-assembly process of $CsPbBr_3$ NSs and photo of wafer-sized thin film on a glass slide (72 $cm^2$). Reproduced from Ref ([57]). Copyright 2019, American Chemical Society. f) RL spectra of X-ray scintillation screen based on $CsPbBr_3$ NSs thin films with varied thickness. The inset shows the PL spectra of fresh and aged samples, showing no significant difference. Reproduced or adapted with permission from Ref ([58]). Copyright 2019, Springer. g) Photograph of $Mn^{2+}:CsPbBr_{1.5}Cl_{1.5}$ PNCs solution at increasing concentration under ultraviolet illumination. Reproduced from Ref ([59]). Copyright 2019, American Chemical Society.

This approach was later extended to other lead halide compositions by Chen *et al.*[12], demonstrating that color-tunable PNCs scintillators can provide a convenient visualization tool for X-ray radiography. Different halide compositions (i.e. using Cl, Br or I) were obtained by controlling the reaction of Cs-oleate with different $PbX_2$ precursors, again using a HI solution method. These oleate-capped PNCs were then cast onto the flexible substrates, allowing rapid multicolor X-ray visualization that was previously inaccessible to bulk scintillators (**Figure 2d**). For low dose irradiation of 5.0 μGy·$s^{-1}$ at 10 kV, the scintillation capability of $CsPbBr_3$ PNCs thin films (thickness=0.1 mm) was comparable to that of high-efficiency CsI:Tl bulk scintillators (thickness=5.0 mm), whereas it was much more favorable (more intense by a factor of 5 or more) than other bulk scintillators, including $PbWO_4$, $YAlO_3$:Ce and $Bi_4Ge_3O_{12}$ (BGO). Solution-processed PNCs prepared by a similar strategy were also used as direct detectors for soft X-rays, showing an impressive detection limit as low as 13 nGy·$s^{-1}$ [12, 35].

Despite their promising advantages, PNCs proved difficult to cast into compact solid films with the required thickness (>100 μm, due to the long penetration depth of high-energy ionizing radiation) and large area sizes for commercially viable applications. Therefore, only small area devices were initially fabricated and tested[36, 38, 60-62]. This was the case until 2019, when Zhang *et al.*[57] reported a modified HI synthesis protocol that yielded high-quality scintillators in both colloidal and solid film forms, designed and self-assembled from highly concentrated solutions of perovskite nanosheets (NSs). The colloidal form of concentrated $CsPbBr_3$ NSs exhibited RL brightness equivalent to the film counterpart (**Figure 2e**) and showed long-term stability under both storage and X-ray exposure conditions. Due to their high concentration and NSs morphology, they could be easily processed in solution and assembled to large area films without cracks up to 72 $cm^2$ (inset of **Figure 2e**), enabling high resolution (< 0.21 mm) X-ray imaging applications. Crucially, the $CsPbBr_3$ colloids exhibited a higher LY (~21000 photons/MeV) than the commercially available Ce:LuAG single crystal scintillator (~18000 photons/MeV). Scintillators based on these NSs showed both strong RL and long-term stability under low-rate X-ray illumination (18 $\mu G \cdot s^{-1}$). Another work with similar $CsPbBr_3$ NSs was reported by Wang *et al.*[58] using a green synthesis approach. They reported a room temperature gram-scale production of $CsPbBr_3$ NSs with minimal solvent usage, saving more than 95% of the solvent for the unit mass NCs production. The perovskite colloid exhibited record stability during long-term storage of up to 8 months (inset of **Figure 2f**), maintaining a PLQY of 63% in the solid state, allowing for crack-free thin films with emission intensity linearly proportional to their thickness, indicating a negligible contribution of the reabsorption effect to their performance (**Figure 2f**). Also on the green synthesis side, a recent work by Erroi *et al.*[63] reported the large scale synthesis of $CsPbBr_3$ PNCs through a modification of the LARP method with the assistance of a turbo-emulsifier that enabled multigram production of PNCs with reusable solvents and recovered precursors and the fabrication of high LY scintillators (up to 5000 photons/MeV) by their direct incorporation into polyacrylate slabs.

While PNCs have bright emission, it is crucial to also achieve large Stokes shifts to suppress the detrimental effects of reabsorption of the scintillation light, especially in the case of highly concentrated composites or in densely packed PNC solids[64-66]. These optical properties are typically achieved by doping $CsPbX_3$ PNCs with dopants such as transition metals (Cu, Mn)[59, 67-77] or rare earths (Lu, Yb, Ce)[78, 79]. In all these configurations, optical reabsorption is minimized by decoupling the dopant emission from the absorption of the perovskite matrix, resulting in a large "apparent" Stokes shift, but at the expense of luminescence lifetime that is typically extended to the microsecond or even millisecond range. Although the relatively poor chemical stability and the intrinsically low concentration that can be achieved by using oleyl chains as ligands limit the efficiency of doped PNCs as scintillators[80, 81], recently Kovalenko and group[59] demonstrated a PNC-based scintillator that simultaneously exhibits high PLQYs (>50%), high particle concentration (>100 mg/mL), and a large Stokes shift of ~1eV by using long-chain zwitterionic ligands in the synthesis of $Mn^{2+}$-doped $CsPb(Br_{0.5}Cl_{0.5})_3$ PNCs (**Figure 2g**). These NCs exhibited more than eight times brighter emission under fast neutron irradiation than their oleyl-capped variants.

Progress in scintillator technology using PNCs is largely dependent on three pillars of development, starting with the interaction of PNCs with ionizing radiation, the stability of these NCs in different environments and the radiation hardness of PNCs. The following section describes development efforts and notable research on these parameters in the quest to develop the optimal material for scintillators. One such approach was presented by Cao *et al.*[82], in which $CsPbBr_3$@$Cs_4PbBr_6$ with emissive $CsPbBr_3$ PNCs embedded in a solid $Cs_4PbBr_6$ host (structure shown in **Figure 3a**) was synthesized using an "emitter-in-matrix" principle and subjected to X-ray sensing and imaging. This design offers several advantages: firstly, the $Cs_4PbBr_6$ matrix not only enhances X-ray attenuation but also significantly improves the stability of the embedded $CsPbBr_3$ PNCs. Secondly, the broadband gap $Cs_4PbBr_6$ makes it transparent to the green emission of the $CsPbBr_3$ PNCs for efficient light output, resulting in low reabsorption losses. Finally, the facile solution synthesis provides us with large scalability, excellent repeatability, low-cost fabrication and high PLQY up to 60%. As a result, the $CsPbBr_3$@$Cs_4PbBr_6$ PNCs exhibit scintillation response to X-rays with excellent linearity and ultrashort decay time, ensuring low-dose sensing and high-quality imaging. A similar approach has been discussed by Xu *et al.*[83], who investigated low-temperature solution-grown $CsPbBr_3$ PNCs embedded in a 0D $Cs_4PbBr_6$ microstructure material with strong green light emission. These PNCs were found to exhibit desirable scintillation properties such as visible emission (531 nm), fast decay time (<10 ns), fine energy resolution (3.0 ± 0.1% at 59.6 keV), high LY (64000 photons/MeV) and long-term stability in air.

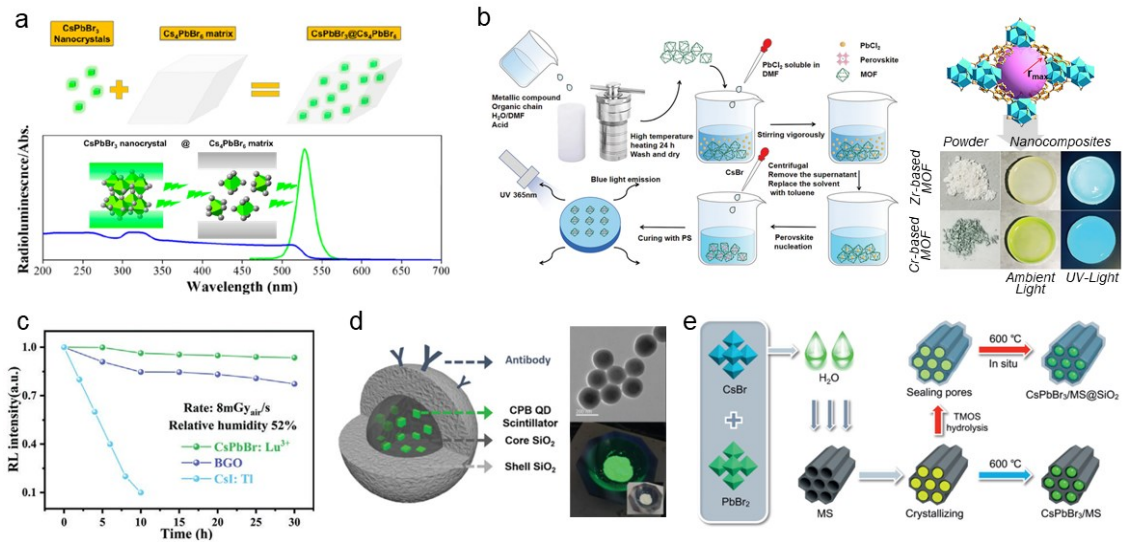

**Figure 3:** a) Schematic approach (top) and absorption/RL spectra of $CsPbBr_3$@ $Cs_4PbBr_6$ PNCs. Reproduced from Ref ([82]). Copyright 2019, American Chemical Society. b) Synthetic strategy and structure of MOF-confined $CsPbCl_2Br$ PNCs. MOF powder and scintillation composites discs obtained by curing the $CsPbCl_2Br$@MIL-101(Cr) and $CsPbCl_2Br$@UiO-67-bpy composites with PS resin under different light sources. Reproduced or adapted with permission from Ref ([84]). Copyright 2022, Royal Society of Chemistry. c) RL intensity as a function of X-ray exposure time for $Lu^{3+}$-doped $CsPbBr_3$ PNCs compared to typical commercial inorganic single crystal scintillators BGO and CsI:Tl. Reproduced or adapted with permission from Ref ([85]). Copyright 2021, Wiley. Bottom-up (d) and top-down (e) approached to prepare silica-coated PNCs starting from colloidal $CsPbBr_3$ PNCs and mesoporous silica nanoparticles, respectively. Reproduced or adapted with permission from Ref ([86]) and ([87]). Copyright 2021, Wiley.

Despite the numerous benefits associated with the utilization of PNCs in scintillation, it is noteworthy to mention that researchers have actively addressed the challenge of low stability associated with PNCs in recent years through the exploration of diverse chemical passivation and confinement methods. One such approach is to explore the idea of confining PNCs in porous metal-organic framework (MOF) matrices, which consist of a metal cluster sublattice coordinated by organic chains acting as framework support and spacers (MOF structure shown in **Figure 3b**). Their specific surface area is large and the pore sizes can be tuned to the PNCs by adjusting the length of the organic chains[88], resulting in a highly stable environment capable of preserving the optical properties of the PNCs under extreme conditions such as water, acid-based solutions and high temperatures[89]. Using this approach, Ren realized scintillators consisting of $CsPbCl_2Br$ PNCs inside a MOF structure[84]. To ensure the uniformity of the PNCs inside the MOF, $PbCl_2$ was first introduced into the aperture of the MOF[90], then the excess $PbCl_2$ was washed out in the solution, and finally CsBr was added to form the composite material $CsPbCl_2Br$@MOF. Long-term stability was achieved by embedding the $CsPbCl_2Br$@MOF composite in a UV-cured polystyrene matrix. The spatial confinement induced by the MOF structure allowed to obtain a scintillator with emission centered at 445 nm and a PL lifetime of 2.16 ns, significantly faster than unconfined PNCs of the same composition. Another interesting approach to improve the stability and radiation hardness of PNCs was reported by Zhang *et al.*[85], who synthesized $CsPbBr_3:Lu^{3+}$ PNCs embedded in a transparent amorphous network structure using a self-limited growth strategy. The $CsPbBr_3$ PNCs with $Lu^{3+}$ ions incorporated in the transparent medium exhibited superior visible light transmittance and significantly improved emission intensity under UV irradiation compared to the counterpart without $Lu^{3+}$. The internal quantum efficiency was found to increase from 25.67% to 65.7% after the introduction of $Lu^{3+}$. Consequently, the optimized RL intensity of the $CsPbBr_3:Lu^{3+}$ was comparable to that of the commercial BGO sample, accompanied by a faster PL decay time of $\tau = 27$ ns and an even higher resistance to irradiation than the control BGO sample (**Figure 3c**). In particular, the $CsPbBr_3:Lu^{3+}$ scintillator showed a strong X-ray absorption efficiency from ~36 to 60 keV, covering the range of common medical digital radiography.

Recently, innovative strategies for the realization of high-quality scintillating $CsPbX_3$ PNCs in impermeable host matrices have also been proposed. The silica coating is a powerful strategy adopted to isolate PNCs from the surrounding environment and to overcome their typical low stability in aqueous solution, which leads to their rapid dissolution and further consequent release of potentially harmful heavy metal ions (e.g. $Pb^{2+}$), an issue to be considered in case of application of PNCs in biological contexts. Different strategies have been pursued to achieve this goal: one example is a bottom-up approach, in which an inorganic robust $SiO_2$ shell was grown around previously synthesized PNCs[86] by rapidly injecting a precursor solution containing the basic catalyst for $SiO_2$ synthesis into a Si precursor solution. This resulted in the simultaneous synthesis of $SiO_2$ NPs and LARP of $CsPbBr_3$ PNCs, which ultimately resulted in the LHP NCs being trapped inside the $SiO_2$ NPs prior to aggregation or decomposition **(Figure 3d)**. As an alternative approach, Li *et al.* obtained a similar structure through a top-down strategy consisting of using a scaffold of silica mesoporous nanoparticles (MSNs) as nucleation hosts for PNCs formation[91-93], which also determined the final size of the PNCs. The MSNs-embedded PNCs were prepared by soaking the bare MSNs in the solution of the precursor salts (CsX and $PbX_2$, where X = Cl, Br or I), followed by drying and heating at 400-800 °C. The presence of a sintering agent, which promotes the collapse of the internal pores of the MSN template, ensured the sealing of the emissive PNCs inside the MSN (**Figure 3e**). The silica-coated PNCs have been successfully applied as biological markers for in vivo X-ray imaging and have been shown to dramatically enhance the production of radical oxygen species for radiotherapy applications[94]. Similarly, an armour-like passivation strategy to synthesize highly efficient and stable $CsPbBr_3$ PNCs by solvothermal method was presented by Yang *et al*[95]. The protective layer formed by hand-in-hand ligands and the electrostatic adsorption formed by ammonium salts with high steric hindrance give the passivated $CsPbBr_3$ PNCs a PLQY=97±3% and a storage stability of up to three months in air and one month in water, preserving 90% of their initial PLQY. Another important result which elucidated the pivotal role of PNCs surfaces engineering has been reported by Zaffalon *et al*[96], who showed that post-synthesis treatment with fluoride sources[97] on PNCs synthetized via HI routes led to a >500% increase in scintillation efficiency reaching LY=8500 photons/MeV, approaching scintillation performance comparable to commercially available single crystal scintillators.

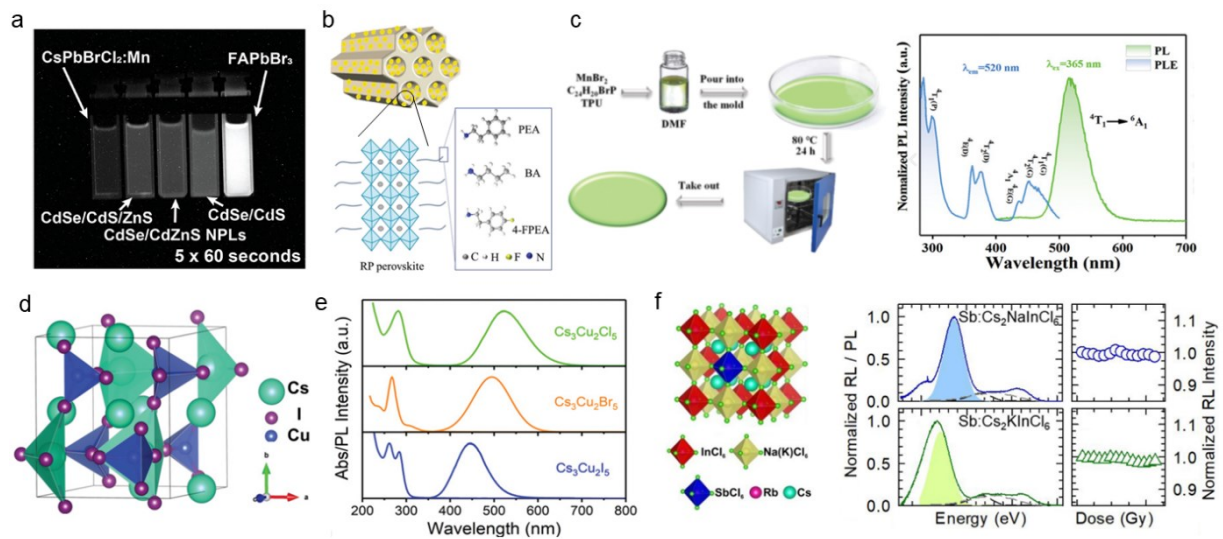

**Figure 4:** a) Radiograph of the five brightest NC emitters, average of five 60-s fast neutron exposures. Reproduced from Ref ([81]). Copyright 2020, American Chemical Society. b) Schematic of RP perovskites comprising three different organic ligands: phenethylamine (PEA), butylamine (BA) and 4-fluorophenethylamine (4-FPEA). Reproduced or adapted with permission from Ref ([98]). Copyright 2023, Wiley. c) Scheme depicting the fabrication of the $(C_{24}H_{20}P)_2MnBr_4$-TPU scintillation film and corresponding PL and PLE spectra. Reproduced or adapted with permission from Ref ([99]). Copyright 2023, Wiley. d) Crystal structure of $Cs_3Cu_2I_5$, with $Cs^+$ cations shown in green, $Cu^+$ ions in blue, and $I^-$ ions purple. Reproduced or adapted with permission from Ref ([100]). Copyright 2021, Wiley. e) Normalized absorption and PL spectra of colloidal solution of $Cs_3Cu_2X_5$ (X = Cl, Br, I) NCs in hexane under UV light (λ = 254 nm). Reproduced or adapted with permission from Ref ([101]). Copyright 2020, Wiley. f) Sketch of the crystal lattices of Sb-Doped "3D" Double Perovskites. Corresponding PL (shaded curves) and RL spectra (solid lines) at room temperature excited using soft X-rays of representative samples 0.7% Sb-doped $Cs_2NaInCl_6$, (top) and 0.9% Sb-doped $Cs_2KInCl_6$. Reproduced from Ref ([102]). Copyright 2021, American Chemical Society.

Off-center approaches for scintillating PNCs: In addition to the approaches discussed above revolving around the engineering of $CsPbX_3$ PNCs, there are several other perovskite nanostructures that have been under investigation for scintillator technologies. Hybrid organic-inorganic compositions obtained by using formamidinium (FA) instead of Cs are particularly interesting for fast neutron detection, with $FaPbBr_3$ PNCs exhibiting the highest LY under neutron irradiation among different LHP NCs compositions, as reported by McCall[81] (**Figure 4a**) and Maddalena[49] in two different papers.

In addition to $APbX_3$-type perovskites, Ruddlesden-Popper (RP)-type perovskites ($A_2B_{n-1}Pb_nX_{3n+1}$; where A and B are long-chain and small organic cations, respectively; and X = Cl, Br, or I is the halide anion) have also attracted attention due to their high stability and strong quantum confinement of excitons. This family of PNCs has been shown to form a natural multiple quantum well system, where efficient ultrafast energy funnelling from small to large $n$ layers (where $n$ is the number of lead halide layered perovskite sheets in an inorganic layer) has been demonstrated in a mixed phase with an efficiency exceeding 85% [103-106]. One such work by Wang *et al.*[98] reported a material design strategy and fabrication processes to achieve RP-type $(A)_2(MA)_{n-1}Pb_nBr_{3n+1}$ PNCs within a mesoporous silica template (**Figure 4b**) exhibiting PLQYs > 99% and long term stability under ambient or ultraviolet irradiation exposure. These RP PNCs not only have good applicative potential for high resolution PL imaging, and waterproof inks but also as stretchable perovskite X-ray scintillators with X-ray imaging with resolution greater than 14 lp·mm$^{-1}$. Another interesting class of materials similar to perovskite structures are two-dimensional (2D) layered metal halides. Such layered crystals of $L_2MX_4$ (where L=monovalent organic cation, M=inorganic divalent cation and X=halide ion) contain multiple quantum well structures with alternating organic-inorganic layers. The inorganic layers of the quantum well consist of corner-sharing $PbX_6^{2-}$ octahedra sandwiched between organic barrier layers. Such a 2D system has four times the exciton binding energy of a corresponding three-dimensional (3D) system[107]. Xia *et al.*[99] used tetraphenylphosphonium as the organic cation in a manganese halide structure and prepared a flexible scintillator in a polyurethane matrix by in situ preparation (**Figure 4c**) to obtain a flexible screen with RL emission six times higher than similar nanocomposites based on $CsPbBr_3$ PNCs. Another example consists of PNCs with phenethylamine cation ($PhePbBr_4$) prepared using a nanostructured glass matrix as a scaffold to obtain size-controlled NCs with PL lifetimes below 3 ns due to the strong quantum confinement effect induced by their reduced size (3-4 nm)[108].

Although all-inorganic lead halide PNCs have exhibited exciting radiation detection properties, the high toxicity and bioaccumulation of lead component in lead halide PNCs is a potential issue that limits their widespread application especially in radio medical diagnostics or radiotherapy. Alternatively, 0D lead-free PNCs also demonstrated to be interesting scintillators. Their high LY and the large intrinsic Stokes-shift due to emission from self-trapped excitons (STE) makes them extremely interesting candidates to be utilized in large area, high sensitivity scintillator devices. Zhou *et al.*[100] reported that $Cs_3Cu_2I_5$ PNCs exhibit a LY four times higher than that of $CsPbBr_3$ PNCs and fabricated an optical fibre panel for obtaining clear projections and computer tomography images (**Figure 4d**). Furthermore, Lian *et al.*[101], showcased the idea by demonstrating its applicability to structures embedding other halides as well, thus realizing solution-processed films with high resolution (**Figure 4e**). Within the family of lead-free PNCs, another promising example of scintillator PNCs was reported by Zhu *et al.*, who demonstrated that Sb-doped double perovskites (DP) systems ($Cs_2B^+B^{3+}X_6$ where $B^+ = Ag^+$, $Na^+$ and $B^{3+} = Sb^{3+}$ and a crystal structure composed of $BX_6$ corner-sharing octahedra surrounded by $A^+$) featured a large Stokes-shift due to particular behavior of $[SbCl_6]$ octahedra which, in absorption, show electronic transitions sensitizing a transparent host matrix and, in emission, act as recombination centers (under both optical and ionizing radiation pumping) undergoing a large structural reorganization depending strongly on the surrounding lattice cage[102] (**Figure 4f**).

- **PNC-nanocomposites for scintillation**

The inherent low stability of PNCs towards moisture, heat, certain radiation and chemicals limits the realization of their application potential[109-112] without further processing. Some of the observed side effects include energy shifts, PL broadening and quenching, and short device lifetime. The low formation energy of PNCs, which also accounts for their straightforward synthesis in colloidal solution, can be attributed to such low stabilities and the aforementioned effects[112-114]. Furthermore, the ionic nature of these materials has been shown to lead to weakening and damage of their ligands when dispersed in polar solvents[112, 115, 116]. In addition, most of the synthetic methods to prepare PNCs use ligands, typically long-chain organic acids and alkylamines, which end up largely serving as preventive measures for PNCs agglomeration[117], rather than surface passivation as in other NC systems. Regarding the embedding of PNCs in polymeric matrices, the direct mixing of NCs and polymers is usually ineffective for the preparation of transparent nanocomposites, since the high specific surface area and surface energy of PNCs, in addition to the strong excluded volume effect of polymers, make the PNCs prone to aggregation. The resulting phase separation within the nanocomposite causes strong scattering-induced transparency losses[118]. A universal way to achieve good transparency is to reduce the aggregation of the NCs. Over the years, several strategies such as surface modification of NCs, in situ polymerization of monomer solutions containing NCs, and in situ growth of inorganic NCs within the polymer matrix have been developed to produce clear nanocomposites[119-123]. However, the high NCs loading of scintillation nanocomposites greatly increase the difficulty of their production, and a careful combination of several strategies, such as surface modification followed by in-situ polymerization, is usually required to accomplish such a task. For their application in scintillator technology, in line with industrial standards, two important parameters need to be further considered: *i*) the ability to fabricate scintillators on a large scale, including large area/volume devices, for which the production of the respective scintillating material in large quantities is paramount. *ii*) the development of device fabrication techniques that are scalable, energy efficient and affordable in terms of raw materials.

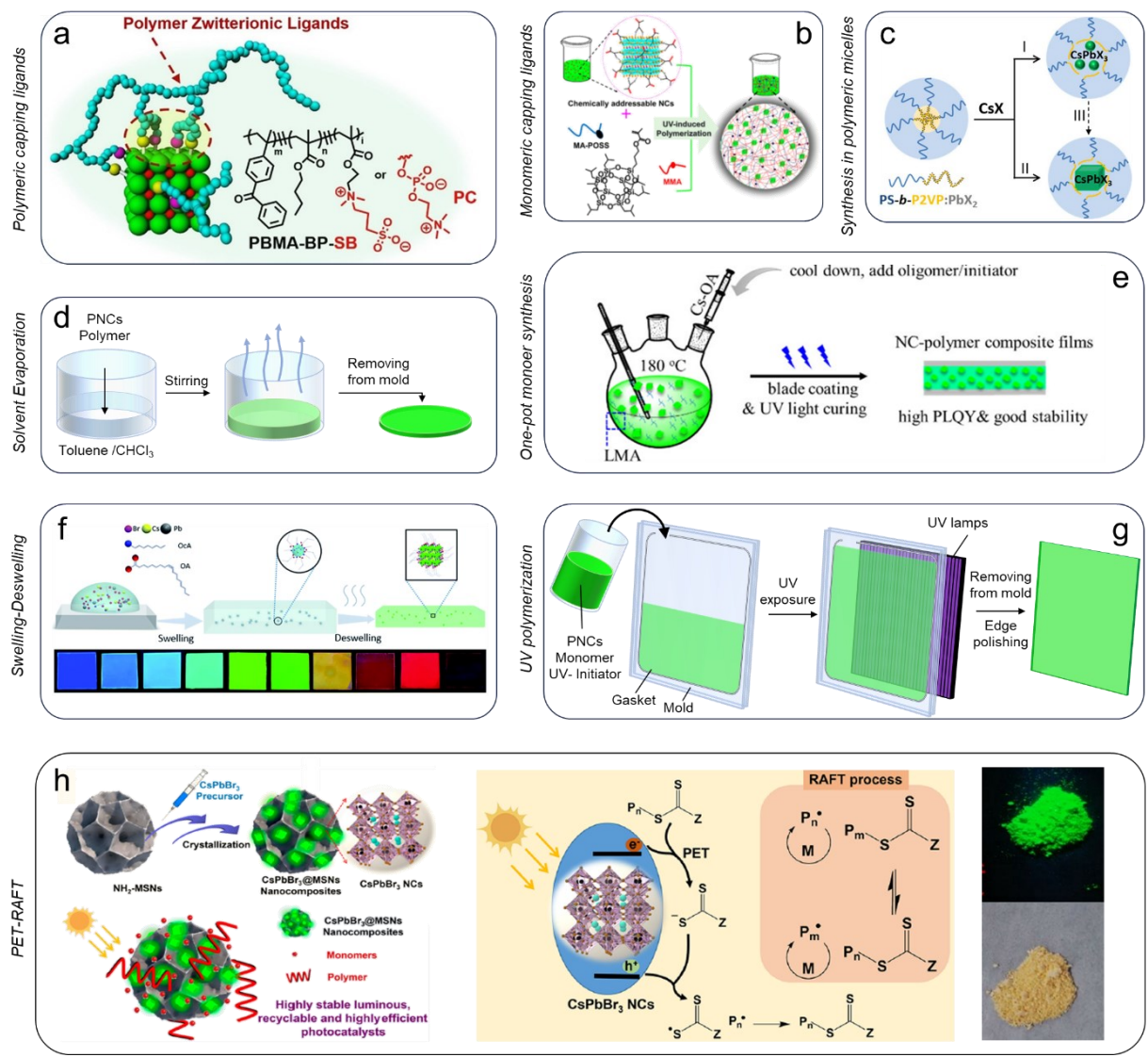

**Figure 5:** Schematic representation of different nanocomposite preparation routes: a) synthesis with polymeric ligands. Reproduced or adapted with permission from Ref ([124]). Copyright 2020, Wiley. b) Synthesis with monomeric ligands. Reproduced from Ref ([125]). Copyright 2018, American Chemistry Society. c) Synthesis in polymer micelles. Reproduced from Ref ([126]). Copyright 2017, American Chemistry Society. d) Solvent evaporation. e) Synthesis in monomeric environment. Reproduced from Ref ([127]). Copyright 2019, American Chemistry Society. f) Swelling/de-swelling technique. Reproduced or adapted with permission from Ref ([128]). Copyright 2020, Royal Society of Chemistry. g) Dispersion and direct UV polymerization. h) PET-RAFT technique. Reproduced from Ref ([129]). Copyright 2022, American Chemistry Society.

The possibility of creating plastic nanocomposites with PNCs has been an area of interest in recent years for the creation of devices for lighting, catalysis or imaging. Among these, scintillation studies with PNCs have justified different approaches to their incorporation as nanocomposites. Various strategies for embedding PNCs in polymeric matrices have been explored in the literature, depending on the specific application. One approach that is particularly useful for biological and/or liquid scintillator applications is the realization of PNCs-loaded micro-powders. These are typically obtained by surface functionalization techniques aimed at embedding few or individual PNCs within polymeric beads. This also offers the possibility of using these powders as building blocks for higher order materials. Alternatively, in-situ synthesis of PNCs directly within the matrix or intercalation of previously synthesised PNCs is used to obtain polymeric coatings. This method allows the production of homogeneous thin nanocomposites (up to hundreds of μm), which are typically used as screens for X-ray imaging, an application that does not require large thicknesses but high precision. Finally, massive nanocomposites for applications in high energy physics and/or rare event detection are typically obtained using bulk polymerization techniques. All these strategies adopted to obtain powders, coating and bulk scintillators are schematically showed in **Figure 5** and described in details in the following sections, with particular emphasis on preserving, or even improving, the scintillating properties of PNCs emitters once embedded in polymeric matrices[63, 130], which represents the most challenging aspect of nanocomposite fabrication.

**Preserving the superior optical properties of PNC when incorporated into polymeric nanocomposites is essential to translating their potential from proof-of-concept to actual, technologically relevant devices.**

Use of monomeric or polymeric capping ligands: This method allows a very thin polymeric layer to grow, making the NCs more affinitive to the polymeric matrix. It is based on the use of monomeric ligands that can be subsequently polymerized by radical polymerization (**Figure 5a** and **b**). Sun et al.[131] introduced a new approach to improve the stability of PNCs through a class of intrinsically cross-linkable and polymerizable. These ligands, which contain a styryl group, provided an opportunity for cross-linking and polymerization crosslinking between the PNCs was achieved by heating with a radical initiator, which effectively improved the stability of the resulting PNCs. They also reported polymer-PNC composites obtained by radical polymerization with excellent transparency and water resistance (**Figure 6b**). Another work by Pan et al.[125] provided an effective and general approach to improve the stability of PNCs via copolymerization with hydrophobic POSS[111] (polyhedral oligomeric silsesquioxane) monomers. POSS-based polymers have been widely used in hybrid materials to induce hydrophobicity, good mechanical and thermal properties[132-134] for the inorganic particles and steric effects, and its low surface energy for hydrophobic passivation layer[132, 133, 135-138]. This approach yielded homogeneous $CsPbX_3$ PNCs with a size of 14-17 nm and PLQY=85% with bound cross-linkable monomers as surface ligands. In addition, the composite could also be dispersed in organic solvents to form readily processable inks, facilitating spin coating, and used in a white down-converting LED.

In the case of polymeric ligands, PNCs are functionalized with oligomeric moieties allowing their self-organization into polymeric micelles, or alternatively the same ligands can actively participate in synthesizing PNCs as capping agents due to their halide content. Using this approach, Kim et al.[124] used polymeric zwitterions to stabilize $CsPbBr_3$ PNCs and nanocomposite films. These PNC-polymer nanocomposites showed considerable resistance to degradation in the presence of polar solvents and benefited from the easy introduction of functionality, which later facilitated patterning by photolithographic methods. Block copolymers containing two or more chemically distinct polymer segments are also widely used as ligands to stabilize [139], providing improved chemical and colloidal stability compared to small molecule ligands [140] and precise control of the spatial arrangement of NCs over different length scales [141-148]. Applying this, Yang et al.[149] reported a facile and general strategy for the preparation of aqueous colloidally stable PS-bpoly(ethyloxide) (PS-b-PEO) grafted MAPbBr$_3$ PNCs (PNC@PS-b-PEO) with excellent long-term stability against water, heat and light. The hydrophobic PS block around the PNCs was reported to provide robust nanoscale encapsulation against water, and the hydrophilic PEO block provided colloidal stability and cytocompatibility in aqueous solutions.

Synthesis of PNCs in polymeric micelles: This approach creates a nano-sized shell of polymeric ligands that bind tightly to a PNC via multidentate coordination groups. The encapsulation process starts by adsorbing one of the PNC precursors into a porous polymer template. This mixture is then dissolved in a solution with the other PNC precursor. Such polymer-coated PNCs are then converted into a nanocomposite by a radical polymerization reaction (see below) or by the solvent evaporation method described above (**Figure 5c**). A detailed demonstration of this approach for insulating PNCs was first shown by Hou et al.[126]. First, a polystyrene-poly(2-vinylpyridine) (PS-b-P2VP) diblock copolymer was added to toluene, where the polymer spontaneously formed core/shell micelles, with the P2VP moiety forming the core and the PS blocks facing the external environment.

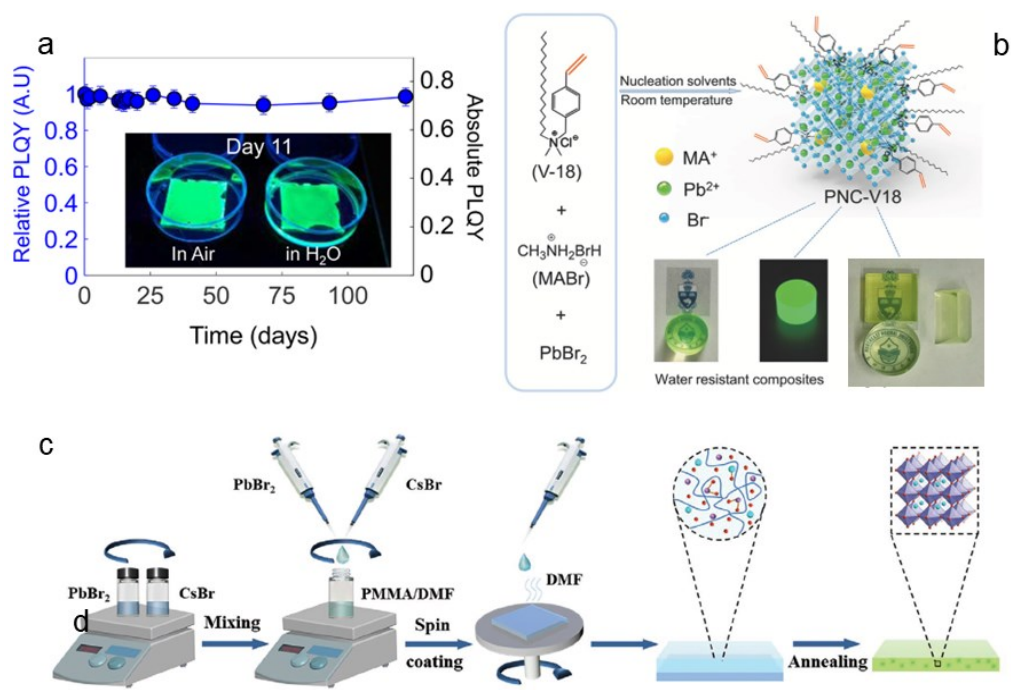

**Figure 6:** a) Relative and absolute quantum yield of 150 μm thick PNCs−polymer composite films after 4 months of water-soaking. Inset: Pictures of polymer composite film after 11 days of soaking in water and in air. Reproduced from Ref ([150]). Copyright 2016, American Chemistry Society. b) Schematic representation of monomeric capping ligand strategy and pictures of nanocomposites evidencing their excellent transparency and water resistance. Reproduced or adapted with permission from Ref ([131]). Copyright 2017, Wiley. c) Schematic representation of the swelling-deswelling procedure to obtain a PNCs-based polymer film. Reproduced or adapted with permission from Ref ([151]). Copyright 2022, Wiley.

Perovskite precursor salts ($PbX_2$ and CsX with X = Cl, Br and I) added to the dispersion accumulate in the cores of the micelles by diffusion. Here the solubility product changes, and the precursors crystallize to form the perovskite phase via an entropically driven process. In this way, the core confines the volume available for perovskite formation, while the shell separates individual reservoirs. The confined growth of the PNCs through the copolymer micelle resulted in the perovskite/polymer core/shell nanostructure, with the formation of strong multidentate bonds effectively passivating the PNC surface. The hydrophobic shell not only acts as a barrier to polar solvents, leading to improved stability in both colloidal and thin film forms, but also provides an additional layer for further surface modification. In a similar study, Hintermayr et al.[152] reported a strategy to employ polymer micelles as nanoreactors for the synthesis of methylammonium lead trihalide PNCs. These PNCs, encapsulated by a polymer shell, showed strong stability against water degradation and halide migration. Thin films prepared from these PNCs showed a 15-fold increase in lifetime compared to unprotected particles under ambient conditions and were reported to survive over 75 days of total immersion in water. The resulting PNC-polymer composites exhibited PLQY=63% and tunability of the emission wavelength throughout the visible range. This line of research was further extended by Hui et al.[153] to different particle sizes and morphologies of perovskite nanostructures. Very recently Zhang et al.[154] demonstrated a solid-state method for obtaining polymeric beads containing $CsPbBr_3$ nanophases by high-speed grinding with PLQY higher than 90%.

Solvent Evaporation method: In this approach, PNCs, whether grown by hot injection or LARP, are purified and redissolved in volatile solvents such as toluene and dichloromethane (**Figure 5d**). Castable blends are then formed by mixing PNCs and polymers in the common solvent, and the nanocomposite is obtained by allowing the solvent to evaporate. This method was demonstrated by Raja et al.[150] in their work where they encapsulated $CsPbBr_3$ PNCs in hydrophobic bulk polymers and reported retention of the PLQY for several months even in aqueous solvent (**Figure 6a**). In this study, three different polymer matrices with unique hydrophobicity were investigated. Due to their ionic crystal nature and dynamic ligand surface coverage, PNCs are very sensitive to the hydrophobicity/polarity of the chosen polymer matrix. Furthermore, the dynamic ligand coverage on these PNCs is susceptible to being lost during encapsulation in a polymer matrix, resulting in electronic, structural and optical losses.

Therefore, for the encapsulation of PNCs in nanocomposites, it is important to consider four aspects: *i)* special care should be taken in the choice of a sufficiently hydrophobic polymer and a compatible solvent, *ii)* use of low temperature routes to avoid thermal damage to the PNCs, *iii)* use polymers containing alkyl side chains similar to the native alkyl chain on the PNC surface and *iv)* perform surface passivation of PNCs to reduce scattering and increase emission yields[155]. This adaptation produces better NC-polymer interfaces than those with less hydrophobic polymers. After polymer encapsulation, several authors reported stable PLQY for months when soaked in water. Moreover, photostability was observed to be greatly enhanced in the polymer-encapsulated PNCs, which could sustain over $10^{10}$ absorption events per PNC prior to photodegradation. This strategy has been used in many explorative studies of the scintillation properties of PNCs, such as the work on scintillation timing by Čuba and co-workers[155-157] described in detail below, and by the Sellin group using mixed-halide $CsPbCl_xBr_{(1-x)}$ PNCs[158] in PMMA and $FAPbBr_3$ PNCs in PMMA and PVT matrixes[159]. Also using the solvent evaporation method, Gandini et al.[15] produced PMMA nanocomposites incorporating both $CsPbBr_3$ PNCs and a large Stokes shift perylene dyad with optical absorption resonant to the PNCs emission, which resulted in sensitized RL free from reabsorption losses.

One-pot monomer synthesis: A more direct approach to prepare PNC nanocomposites is to synthesize PNCs directly in the matrix, with the advantage of avoiding post-synthesis strategies, reducing both material waste and optical performance loss (**Figure 5e**). In this case, a solution of high-boiling monomer is used as a reaction solvent for the preparation of PNCs by hot injection, as demonstrated by Tong et al.[127], who reported a post-processing-free protocol (i.e. without centrifugation, separation and dispersion process) for the synthesis of $CsPbBr_3$ PNCs in LMA monomer, which was then directly combined with oligomers and initiators by UV polymerization to prepare the high-quality PNC-polymer composite films. These films exhibited PLQY=85-90% and FWHM as narrow as 20 nm. Yang et al.[160] in an independent study developed a facile method to synthesize $CsPbBr_3$ in poly-diphenylvinylphosphine-styrene (P-DPEP) composites by thermal polymerization on the basis of a one-pot hot injection process. The resulting composites exhibited an impressive PLQY=90% using DPEP as the surface ligand and polymerization centre. Importantly, these composites exhibited significantly improved long-term stability in the presence of water, methanol and UV light irradiation compared to pure $CsPbBr_3$ PNCs.

Swelling-Deswelling method (SDM): Similar to one-pot monomer synthesis, SDM takes advantage of the presence of polymer chains to perform PNCs synthesis (**Figure 5f**). PNCs precursors are added to a high-viscosity solution consisting of polymer dissolved in an organic solvent (swelling step). Once the mixture is poured onto a substrate, the solvent is removed by heating (deswelling step) and PNCs are progressively formed inside the matrix, while aggregation is avoided thanks to the extremely high viscosity of the mixture[161]. SDM has already been used for drug delivery applications, introducing medicines as solutes into polymer matrices and allowing their controlled release[162, 163]. Wong et al.[164] in 2016 exploited this phenomenon to synthesize ultra-stable, highly luminescent organic-inorganic perovskite polymer (OIP) composite films. They worked on the assumption that OIP precursors can be introduced as solutes into polymer matrices through the solvent-induced polymer swelling process. When the solvent was driven out of the polymer matrix by heating, the OIP precursors were left within to react and form high quality, well dispersed OIP NCs (**Figure 6c**). Meanwhile, the polymer matrix deswelled, shrunk back and formed a coherent barrier layer around the PNCs, protecting them from water, oxygen or heat, enabling the PNCs to withstand boiling water treatment for 30 minutes or two months in water at room temperature with an efficiency loss of less than 7%. However, this strategy could only produce bromide-based perovskites with good performance, while other halide compositions had low emission and were still vulnerable to external agents. Researchers from the same group in 2020 reported a ligand-assisted SDM (LA-SDM) method to synthesize PNCs inside polymer matrices with high stability, wide colour tunability and good colour purity, and demonstrated it as a facile strategy to be applied to diverse perovskite precursor compounds, various polymer substrates and different processes[128]. In this work, the authors modified the protocol where ligands were applied in SDM to help with micelle formation during the swelling process, to confine small PNCs crystallization, and subsequently act as coordinators and stabilizers after crystallization. This resulted in PLQY > 70% for composites with a wide range of optical densities (0.05–0.77) without concentration quenching effects. Composites with emission color ranging from the blue to the near infrared were obtained by tuning the halide compositions in the precursor solution.

Direct Radical Polymerization: Despite the wide variety of approaches to produce nanocomposite scintillators, most of them have intrinsic limitations for producing nanocomposites suitable for real applications, mainly due to dimensional constraints and difficult upscaling. One of the few strategies that allows to obtain nanocomposites with potentially unlimited dimensions and volumes is direct radical polymerization.

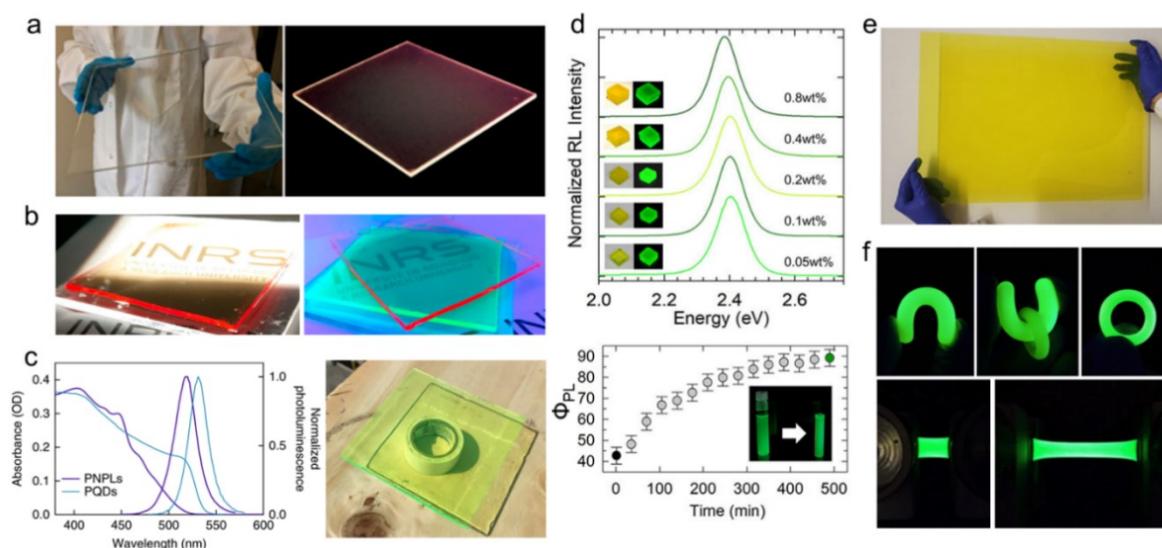

**Figure 7**: a) Pictures of a LSC comprising Mn/CsPbCl$_3$ PNCs under ambient (left) and UV illumination (365 nm, right). Reproduced from Ref ([69]). Copyright 2017, American Chemistry Society. b) Pictures of LSC embedding CsPb(Br$_{0.2}$I$_{0.8}$)$_3$ PNCs under solar (left picture) and UV (right picture) illumination. Reproduced or adapted with permission from Ref ([166]). Copyright 2017, Elsevier. c) PL and absorption spectra of layered perovskite PNCs (R$_2$(MA)$_{n-1}$Pb$_n$Br$_{3n+1}$, with R=hexylammonium) embedded in a PMMA-based LSC. Reproduced or adapted with permission from Ref ([172]). Copyright 2019, Springer. d) RL spectra of nanocomposite with increasing CsPbBr$_3$ loading (top) and PLQY (bottom) during the UV-polymerization process of CsPbBr$_3$-PMMA/PLMA nanocomposite. e) Picture of large-area (60x60 cm$^2$) scintillator based on CsPbBr$_3$ PNCs fabricated via direct photopolymerization. Panels d and e are reproduced from Ref ([63]). Copyright 2023, American Chemistry Society. f) Pictures of the self-standing CsPbBr$_3$−PBMA rubbery cylinders during flexing (top) and stretching (bottom) test taken under UV illumination. Reproduced with permission from Ref ([173]). Copyright 2018, American Chemistry Society.

This approach has been used to realize large-area polymeric luminescent solar concentrators (LSC)[165, 166], in which chalcogenide NCs and/or organic dyes are embedded[65, 66, 167-171]. Typically, a monomer solution containing the PNCs and radical initiator is poured into a mould (**Figure 5g**) and thermal or optical energy is provided to initiate radical generation. Since high temperatures typically damage PNCs[165, 174], UV photopolymerization is generally preferred, using initiators such as diphenyl (2,4,6-trimethylbenzoyl) phosphine oxide or 2,2-dimethoxy-2-phenylacetophenone. This approach has been used in several LSC works, such as the study by Meinardi *et al.* who fabricated large bulk composites by photopolymerizing polymethyl/lauryl methacrylate (PMMA/PLMA) embedding manganese-doped $CsPbCl_3$ PNCs[69] (**Figure 7a**), Zhao *et al.*[166] who used mixed halide $CsPbX_3$ PNCs (**Figure 7b**) or Wei *et al.*[172] using layered PNCs (**Figure 7c**). If the polymerization is not aggressive but timed appropriately, a beneficial effect on the surface of the PNCs is observed, improving their optical properties, as reported by Erroi *et al.*[63] (**Figure 7d**) who produced large sized nanocomposite scintillators embedding $CsPbBr_3$ PNCs (**Figure 7e**) embedded into a PMMA/PLMA copolymer. Adopting the same strategy with a different monomer selection allows to further extend the mechanical properties of the final nanocomposite. A notable example is provided by Xin *et al.*[173] who used a blend of poly(butyl methacrylate) (PBMA) monomer and $CsPbBr_3$ PNCs to fabricate flexible and stretchable nanocomposites (**Figure 7f**). However, UV photopolymerization, although simple and scalable, has intrinsic limitations in the preparation of highly loaded nanocomposites because the PNCs absorb most of the UV light required to activate the radical initiator, resulting in inhomogeneous or incomplete polymerization of bulk composites. To overcome this limitation, strategies such as photoinduced electron/energy transfer reversible addition-fragmentation chain transfer (PET-RAFT) offer a potentially valid approach. On the thermal polymerization front, there are currently no examples in the literature of LHP NCs that have retained their original optical properties of PNCs in solution, although some examples of thermal quenching suppression have been reported, mostly for PNC-based LEDs[97, 175-177], which may provide a suitable fabrication strategy also for thermally polymerized nanocomposites.

Photoinduced Electron/Energy Transfer Reversible Addition-Fragmentation Chain-Transfer (PET-RAFT): The PET-RAFT polymerization was introduced in 2014 as a strategy to accurately control the polymer chain distribution and the polymerization kinetics[178]. This approach is based on the use of photocatalysts which, once photo-excited via energy or electron transfer from an optically active species, directly activate the RAFT agent, which undergoes fragmentation and generates the initiating radical species (**Figure 5h**). In general, the photo-initiator can be purified or it could be left in the polymer, giving a perfectly dispersed filler in the composite once the solvent is removed[179]. This technique was already implemented using different types of nanoparticles such as $CsPbBr_3$ PNCs[180, 181] and PNC embedded into mesoporous silica nanoparticles[129]. Importantly for nanoscintillators application, PET-RAFT polymerization allows to obtain nanocomposite with high optical quality even at high concentrations of dissolved species, since the polymeric matrix is grown from the surface of the PNCs, progressively separating them, and preserving their dispersion. Moreover, the livingness polymerization character of this strategy, which involves the absence of chain termination/transfer reactions and an initiation the rate much larger than the rate of chain propagation, results in a constant growth rate than seen in traditional polymerization and, ultimately in a very low polydispersity index[182], which directly affects the optical properties of the composite.

- **Scintillation properties of PNC nanoscintillators**

<u>Impact of defects on the PNC scintillation.</u> In parallel with the synthesis of scintillating PNCs and related plastic nanocomposites, studies have been carried out to understand the fundamental aspects of the scintillation processes in this class of materials. The efficiency of a scintillator is affected by the presence of defects acting as trapping sites which can temporarily capture free charge carriers generated upon the primary interaction, either delaying their radiative recombination or decreasing the overall performance. Although synthesis strategies are continuously improved, a certain number of defects at the atomic scale is very common even for materials synthesized in highly controlled conditions. Therefore, the understanding of the role of defects in the scintillation mechanism becomes essential in the science of scintillators[160, 183-185] and to optimize the LY and timing performance. The scintillation efficiency is expressed by the LY, which is a comprehensive measure quantifying the number of photons emitted per deposited MeV. Different measurement techniques result in LY values spanning several orders of magnitude: for example, the LY of $CsPbBr_3$ PNCs reported in the literature ranges from 2500 to 15000 photons/MeV[82, 85, 96, 186-189]. The optimal measurement conditions are still being debated, and an important distinction should be made between *intrinsic* LY, which is an intrinsic property of the scintillator, and *extrinsic* LY, also called light output, which is the measured physical quantity. This distinction is similar to the internal and external quantum efficiencies (IQE and EQE, respectively) of LEDs: IQE is the number of photons generated inside an LED structure, whereas the EQE quantifies the electroluminescence photons outcoupled outside the device per injected carrier. Many factors contribute to the measured LY, including the ability of the material to interact efficiently with the incident ionizing radiation, the transport efficiency of the generated carriers to the emission centres, their PLQY, and the light outcoupling. While transport and PLQY are determined by the presence of defects trapping the generated excitons and are discussed in the next section, the probability of the scintillator interacting with the ionizing radiation depends on the composition and geometry of the material, while the light collection efficiency is strongly influenced by self-absorption, internal light guiding, multiple reflections and scattering losses, leading to significant variability in the measured LY. In the literature, the measurement of the LY of an inorganic single crystalline scintillator is uniquely defined by the observation of the photoelectric peak in the pulse height spectrum, measured by excitation with a $^{137}Cs$ gamma source and with an acquisition shaping time of 1 μs. In the case of scintillating NCs or nanocomposites with low density, the photoelectric peak is difficult to observe, and a relative method is used for the LY measurement, which can follow three alternative approaches: (*i*) the RL intensity corrected for the fraction of absorbed X-rays evaluated by monitoring the variation of the response of a reference scintillator[190], (*ii*) a side-by-side comparison of the RL intensity with a reference scintillator in identical geometrical conditions, and (*iii*) the RL intensity corrected for the X-ray mass attenuation coefficient and the sample thickness, which overestimates the LY by neglecting the self-absorption and the light outcoupling factor. In **Table 1** we report LY values of LHP-PNCs based scintillators produced and processed following various synthesis and fabrication approaches, in comparison to scintillator crystals.

| Feature | Composition | Sample Form | Scintillation Time (ns) | LY (ph MeV$^{-1}$) | LY measure technique | Ref. |
|---|---|---|---|---|---|---|
| *Lead Halide Perovskite Nanocrystals* | CsPbBr$_3$@Cs$_4$PbBr$_6$ | Thin film | -- | 6000 | Reference scintillator (CsI:Tl) | 82 |
| | CsPbBr$_3$ NCs | PMMA nanocomposite | -- | 166000 | Theoretical limit calculated according to ref. | 28 |
| | CsPbBr$_3$ NCs @ BaF$_2$ | Heterostructure | 11 | 6300 | Photoelectric peak excited with $^{137}$Cs | 188 |
| | CsPbBr$_3$ NCs in glass-ceramic matrix | Glass matrix | -- | 4100 | Reference scintillator (BGO) | 85 |
| | CsPbBr$_3$ NCs | Octane solution | -- | 2300 | Reference liquid scintillator (PPO, POPOP in toluene) | 187 |
| | CsPbBr$_3$/Cs$_4$PbBr$_6$ NCs | NC powders | -- | 3600 | Reference scintillator (LYSO) | 186 |
| | CsPbBr$_3$/Cs$_4$PbBr$_6$ NCs | NC powders | -- | 64000 | Reference scintillator (NaI:Tl, CsI:Na) | 83 |
| | CsPbBr$_3$/CsPb$_2$Br$_5$ NCs | Thin film | -- | 19200 | Reference scintillator (LuAG:Ce) | 191 |
| | CsPbBr$_3$/CsPb$_2$Br$_5$ NCs | Water solution | -- | ~3000 | Reference scintillator (PPO in LAB (2 gL$^{-1}$)) + X-ray attenuation correction | 192 |
| | CsPbBr$_3$ NCs + dye | Solution | -- | 16700 | Reference scintillator (EJ-301) | 193 |
| | CsPbBr$_3$ NCs | NCs in anodized Al oxide | -- | 11100 | Reference scintillator (LYSO) | 91 |
| | CsPbBr$_3$ NSs | Thin film | -- | 21000 | Reference scintillator (Ce:LuAG) | 57 |
| | CsPbBr$_3$ NWs | NWs in anodized Al oxide | -- | 13200 | Reference scintillator (YAG:Ce) + X-ray attenuation correction | 194 |
| | CsPbBr$_3$ NCs | -- | -- | 24000 | Photoelectric peak excited with $^{137}$Cs | 195 |
| | MAPbBr$_3$ NCs | Thin film | -- | 14600 | Reference scintillator (NaI:Tl) | 189 |
| | CsPbBr$_3$:F NCs | Powders | -- | 8500 | Reference scintillator (BGO powders) | 96 |
| | CsPbBr$_3$ NCs | Solution | -- | ~8000 | Reference liquid scintillator (EJ-305) | 196 |
| | CsPbBr$_3$ NCs | Film | -- | 21000 | Reference scintillator (Ce:LuAG) | 197 |
| *Polymeric Nanocomposites* | CsPbBr$_3$ NCs | PMMA/PLMA nanocomposite | 4.1 | 4800 | Fraction of absorbed X-rays | 63 |
| | CsPbBr$_3$ NCs | PMMA nanocomposite | -- | 15800 | Reference scintillator (BGO, CsI:Tl) | 151 |
| | CsPbBr$_3$ NCs + dye | PMMA nanocomposite | -- | 9000 | Reference scintillator (LYSO) | 15 |
| *Metal halides* | Cs$_3$Cu$_2$I$_5$ NCs | PDMS nanocomposite | -- | 48800 | Reference Scintillator (LYSO:Ce) | 67 |
| | CsPbCl$_3$:Yb | Powders | -- | 102000 | Fraction of absorbed X-rays | 78 |
| | Cs$_3$Cu$_2$I$_5$ NCs | Thin Film | -- | ~80000 | Reference Scintillator (BGO) | 101 |
| | Gua$_3$SbCl$_6$ | Powders | -- | 2000 | Photoelectric peak excited with $^{57}$Fe | 198 |
| *Ultra-fast scintillators* | CsI | Single Crystal | ~6 | 3200 | Photoelectric peak excited with $^{22}$Na | 199 |
| | BaF$_2$ | Single Crystal | 0.6-0.8 | 1600 | Photoelectric peak excited with $^{137}$Cs | 200 201 |

| | | | | | |
|---|---|---|---|---|---|
| *High LY Scintillator crystals* | γ-CuI + LiI | Single Crystal | <0.45 | 3800 | Reference scintillator (BaF$_2$) | 202 203 |
| | LYSO:Ce | Single Crystal | 41 | ~33000 | -- | 201, 204 |
| | CsI(Tl) | Single Crystal | 1000 | 54000 | -- | 205 |
| | BGO | Single Crystal | 300 | 8500 | -- | 204 |
| | LuAG:Ce | Single Crystal | ~60 | ~18000 | Photoelectric peak | 206 207 |

**Table 1:** The most significant results on perovskite-based nanoscintillators and archetype bulk inorganic scintillators are reported. An extended comparison including also direct radiation detection and is presented in ref. 208.

Crucially, these parameters are closely related to the presence of shallow and deep trap states that introduce detrimental long-term scintillation contributions (in the timescale of several μs or more), also referred to as *afterglow*, due to the thermal release of carriers, whose probability is strictly related to the energy of the traps involved.

An understanding of the role of defects in the scintillation mechanism is essential for the optimization of both the light yield and the timing performance.

In this respect, Rodà et al.[209] and Maddalena et al.[49] reported the first examples of thermally stimulated luminescence (TSL) and afterglow (AG) investigations of PNCs, as powerful tools to access the dynamics of carrier release from trap states, thus expanding the general understanding of loss mechanisms at the nanoscale. For these reasons, it is instructive to introduce these techniques in more detail. A typical TSL experiment consists of two steps, as schematized in **Figure 8a**. First, the sample is exposed to a prolonged irradiation up to a few Gy of delivered dose at the lowest temperature achievable (typically cryogenic temperature) to populate the available trap states that are stable. Then, the irradiation is interrupted, and the sample is gradually heated at a constant rate while monitoring the emission due to carrier release. The temperature increase makes traps unstable and provides the trapped carriers with the appropriate thermal energy to be freed and again available for recombination. The recorded TSL emission, due to radiative recombination of carriers following detrapping, as a function of temperature is called *glow curve*. The emergence of TSL peaks in glow curves is the typical signature of thermally activated detrapping processes. Specifically, the rise of the TSL corresponds to the onset of carrier release from a trap. Once a significant portion of the traps has been emptied, the TSL signal starts to decrease, leading to a typical peak feature in the glow curve. The study of the energetics of a specific TSL peak allows to evaluate the parameters of the related trap such as its energy depth and the escape frequency. On the other hand, in isothermal AG measurements, the trap sites are populated in an analogous manner, but the sample is kept at a constant temperature throughout the light collection as a function of time. Since temperature and time are linearly correlated in TSL experiments, from the comparison of AG and TSL results it is possible to decouple the thermally activated processes from the athermal release of trapped carriers that may occur via tunnelling between isoenergetic and spatially close states. These experiments, together with classical time-resolved temperature-controlled photoluminescence techniques, have been used by Rodà et al. to study the scintillation properties of CsPbBr$_3$ PNCs of different dimensionality (namely nanocubes, nanowires and nanosheets up to the bulk single-crystal limit) and to elucidate the role of surface defects versus bulk lattice defects[209]. These studies showed that the halide vacancies at the surface introduce a manifold of shallow defects in thermal equilibrium with the band edge exciton and are responsible for the broadening of the emission spectra and a slow luminescence tail of ~3 μs.

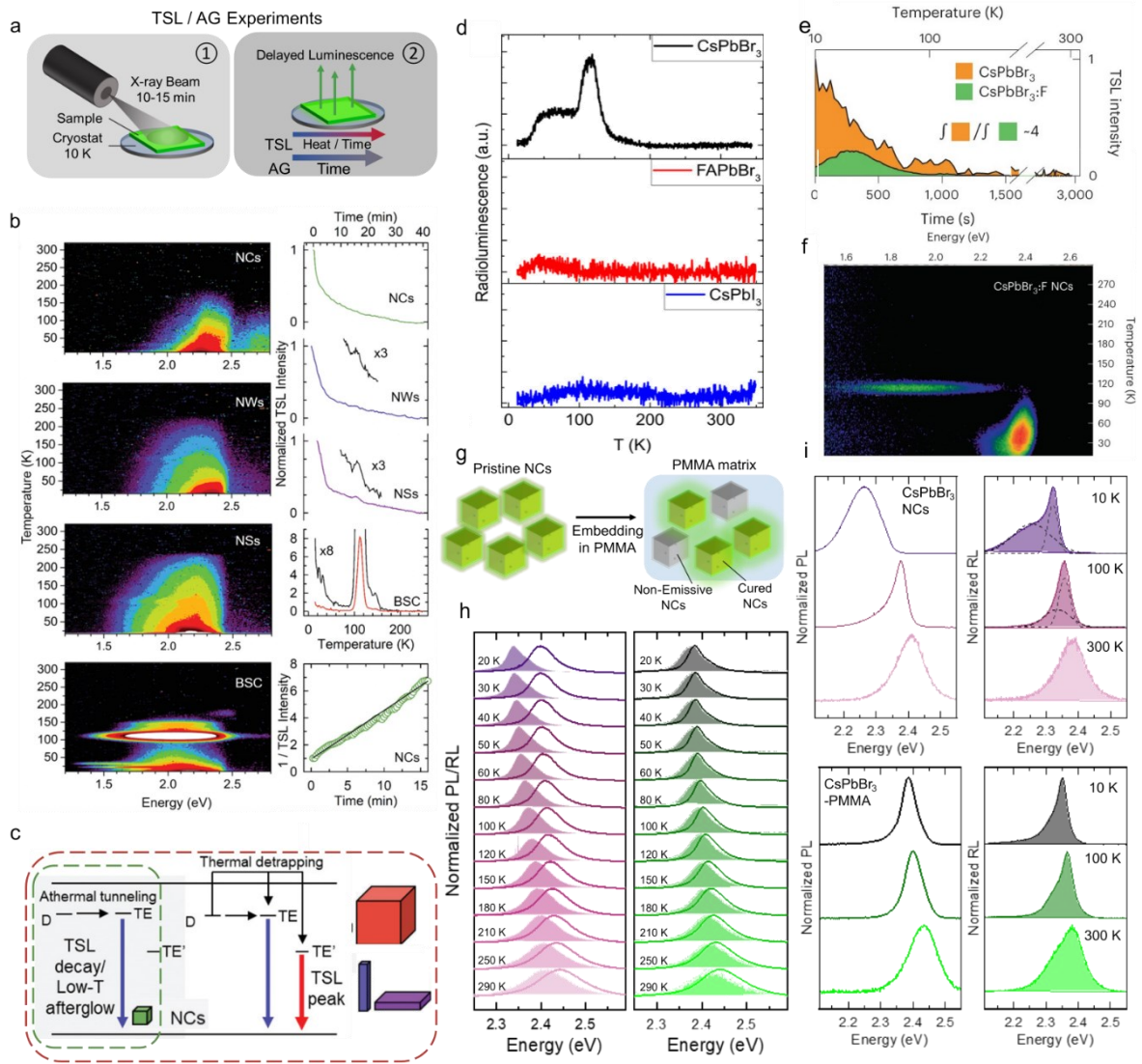

**Figure 8**: a) Schematic depiction of thermally stimulated luminescence and afterglow experiments. b) Contour plots of the spectrally resolved TSL intensity as a function of temperature in the 10–320 K range of CsPBr$_3$ nanostructures (nanocubes, nanowires, nanosheets) and bulk single crystal. In the right top pannel the TSL integrated intensities (glow curve) of all investigated samples normalized for their initial value at $T$ = 10 K. The black insets in are a magnification of the glow curve to highlight the TSL peak of NWs and NSs and the initial monotonic decay of the BSC. In the right bottom panel the inverse of the TSL intensity of CsPbBr$_3$ PNCs as a function of the time-delay after the suppression of irradiation. The black line is the result of the fitting to a power law function I∝(t-t$_0$)$^{-1}$. c) Schematic depiction of the detrapping mechanisms observed in TSL and AG experiments. Panels b and c are reproduced or adapted with permission from Ref ($^{209}$). Copyright 2021, Wiley. d) TSL integrated intensities from low-energy trap states after X-ray excitation for CsPbBr$_3$ (top), FAPbBr$_3$ (centre), and CsPbI$_3$ (bottom) PNCs. Reproduced from Ref ($^{49}$). Copyright 2021, American Chemistry Society. e) Comparison between the TSL intensity of standard and fluorinated CsPbBr$_3$ PNCs corrected for the respective LY values at different temperatures. The integral over temperature (time) of each curve is proportional to the respective trap concentration. f) Same as 'b' for resurfaced CsPbBr$_3$:F PNCs. In the top panel the normalized TSL spectrum integrated between 10 and 80 K (orange line) together with the RL spectrum at $T$ = 10 K (dashed black line). Panels e and f are reproduced or adapted with permission from Ref ($^{96}$). Copyright 2022, Springer. g) PL (solid line) and RL (shaded area) spectra of CsPbBr$_3$ NCs as a function of temperature from 290 to 20 K (from bottom to top). h) Schematic representation of the dual effect of embedding PNCs in PMMA. i) PL (left panel) and RL (right panel) spectra at 10, 100, and 300 K of CsPbBr$_3$ NCs with high defect density (top) and PMMA

nanocomposite (bottom). Dashed lines are the excitonic and defect contributions to the RL. Panels g-i are reproduced from Ref ([210]). Copyright 2024, American Chemistry Society.

As a result, the monotonically decreasing TSL trend shown in **Figure 8b** for the PNCs is due to the release of carriers trapped in shallow defects by athermal tunnelling between spatially correlated states. On the other hand, for bulk systems, the TSL glow curve is dominated by a sharp peak suggesting that the thermal release of carriers occurs from a deep trap state in the gap which can be emptied when the necessary thermal budget is provided according to the scheme in **Figure 8c**. Crucially, the same TSL peak is also observed in the PNCs and intensifies as the dimensionality of the system approaches that of the bulk crystal, suggesting that deep traps are associated with internal lattice defects. These observations highlight the superior potential of PNCs compared to their bulk counterparts, where it is necessary to suppress both core and surface lattice defects. Similarly, in **Figure 8d** Maddalena *et al.*[49] found that $FAPbBr_3$, $CsPbBr_3$, and $CsPbI_3$ PNCs have rather deep trap levels between 100 and 550 meV, ascribed to both surface traps and defects within the PNCs, and that $CsPbBr_3$ shows the most intense low temperature afterglow signal and the highest trap concentration. However, no afterglow at room temperature was ever observed. In addition, Xie *et al.*[211] observed low temperature afterglow in 2D perovskites and correlated it to TSL glow peaks due to traps with energy depths in the range from 20 to 250 meV, ascribed to transient exciton-lattice coupling and, not exclusively, to permanent impurity defects. Crucially, although the "defect tolerance" of $CsPbBr_3$ PNCs is known, both first-principles[212] and atomistic approaches with DFT techniques[213] have clarified that halogen vacancies on the particle surface are the thermodynamically most favoured defect responsible for the shallow defects previously discussed. In this context, Zaffalon *et al.*[96] have recently investigated the effects of surface reconstruction on the scintillation properties of $CsPbBr_3$ PNCs using optical spectroscopy and TSL/AG techniques, elucidating the role of surface and lattice defects on scintillation efficiency. Both RL (X-ray, 20 keV) and cathodoluminescence (electron beam, 20 kV) in standard $CsPbBr_3$ PNCs showed a shallow defect character. Surface treatment with fluorine precursors saturated the Br vacancies at the PNC surface, raised the PLQY to over 90% and increased the LY by 500% compared to standard PNCs, restoring the band-edge character of the RL. In fluorine-rich $CsPbBr_3$:F PNCs, the reduced density of surface defects accounted for an overall ~75% decrease in TSL intensity (**Figure 8e**) and allowed to detect the underlying signal of thermal release from deep red-emitting traps (**Figure 8f**). These observations closely resembled the TSL behaviour of $CsPbBr_3$ bulk counterparts in ref[209] and shown in **Figure 8b** where the same TSL peak has been attributed to crystal lattice defects. These results, which highlight the close relationship between surface defect density and RL efficiency, provide a spectroscopic platform for the development of novel nanoscintillators that are competitive or even superior to commercial BGO or $BaF_2$ crystals[214]. A similar effect to fluorine passivation was observed by Cova[210] et al. in $CsPbBr_3$ PNCs embedded in PMMA nanocomposites, where the passivation of uncoordinated Pb sites on the NCs surfaces by the electron-rich acrylic groups of PMMA restored the band-edge character of RL, as demonstrated by the disappearance of the pronounced redshift of the RL of bare NCs from the corresponding PL, ascribed to the emission from a manifold of shallow defects (**Figure 8g**). This work also clarified the dual effect of polyacrylates on the scintillation efficiency of CsPbBr$_3$ PNCs, as sketched in **Figure 8h**: the degradation of a sub-fraction of the NCs ensemble is counterbalanced by the passivation of electron-poor surface defects of the remaining NCs fraction. This was particularly evident for NCs with high defect density because the strong surface passivation effect of PMMA dramatically enhanced the emission efficiency by greatly suppressing defect-related contributions to scintillation, as shown in **Figure 8i**, where the PL and RL spectrum of the NCs was largely dominated by a low energy emission due to trapped excitons. In addition to defects studies, the nature of the scintillation process and concomitant dynamic loss channels have been recently investigated by means of time-resolved RL experiments, which are discussed in detail in the following section dedicated to the scintillation timing performance, as kinetic processes play a key role in the potential application of PNC scintillators in fast-timing technologies.

<u>Radiation Hardness of PNCs.</u> An important feature in scintillators is the ability to withstand high doses of ionizing radiation, also known as radiation hardness. Lead halides have already been studied in crystalline form for aerospace applications, demonstrating resistance to damage that is largely superior to crystalline silicon[215-219]. At the nanoscale, however, most studies have been devoted to X-ray imaging applications using thin films of PNCs as scintillating screens to be coupled with conventional optical detector systems (CCD, CMOS). In this field, especially for medical imaging applications, radiation levels are necessarily low, so the focus is on optimizing response linearity within the $\mu Gy \cdot s^{-1}$ to $mGy \cdot s^{-1}$ dose rates - to increase image contrast and spatial resolution - rather than radiation hardness. As a result, radiation stability studies have been limited to a few tens of Gy [192, 196, 220, 221] (**Figure 9a, b**), sometimes indicating a certain degree of instability even after a few Gy[85, 196]. For energy conversion schemes the doses involved are typically higher than X-ray imaging, motivating the development of PNCs with high stability in the kGy range. This has been achieved by Yang *et al.*[95] on specifically surface treated CsPbBr$_3$ PNCs, demonstrating RL intensity stability in PNCs up to 42 kGy dose of $^{60}$Co γ-rays (at a dose rate of 0.45 kGy·h$^{-1}$). A different aspect of radiation hardness is that of high-energy particle colliders and nuclear reactors, where the doses involved are up to 5 or 6 orders of magnitude higher than in X-ray imaging, and typically exceed the extreme dose of 1 MGy[222, 223]. To date, only a few studies have directly addressed radiation hardness in LHP NCs above 10$^5$ Gy, mainly from Brovelli and co-workers[63, 96, 210] and from Ostrikov and co-workers[224, 225] using $^{60}$Co γ-rays. In particular, it has been shown that radiation hardness is an intrinsic property of LHP PNCs, regardless of whether they are synthesized by hot injection[96] or by LARP techniques[63, 210], and that this property is retained when embedded in polyacrylate nanocomposites (**Figure 9c, d**) obtained by solvent evaporation or by UV-initiated free-radical polymerization (**Figure 9e**).

> **Extreme radiation hardness is an intrinsic property of LHP PNCs, whether synthesized by hot injection or LARP techniques, and this property is retained when embedded in polymeric nanocomposites.**

Similarly, the radiation hardness of PNCs is independent of resurfacing techniques; the direct comparison of standard and fluorine-resurfaced CsPbBr$_3$ PNCs in **Figure 9f, g** shows that exposure up to 1 MGy of gamma rays does not induce further defect formation, essentially preserving the original spectral properties (optical, scintillation and TSL) independently of their passivation or surface defectiveness[96]. On the other hand, the optical and scintillation performances are strongly dependent on the synthetic routes for the PNCs preparation, retaining both PLQY and LY values both for pristine and F-treated CsPbBr$_3$ PNCs (**Figure 9h**) up to 1 MGy dose of $^{60}$Co γ-rays. More recently also Ostrikov and co-workers[224, 225] reported on $^{60}$Co γ-rays stability of CsPbBr$_3$ PNCs showing 90% retention of RL intensity up to 319 kGy (**Figure 9i**).

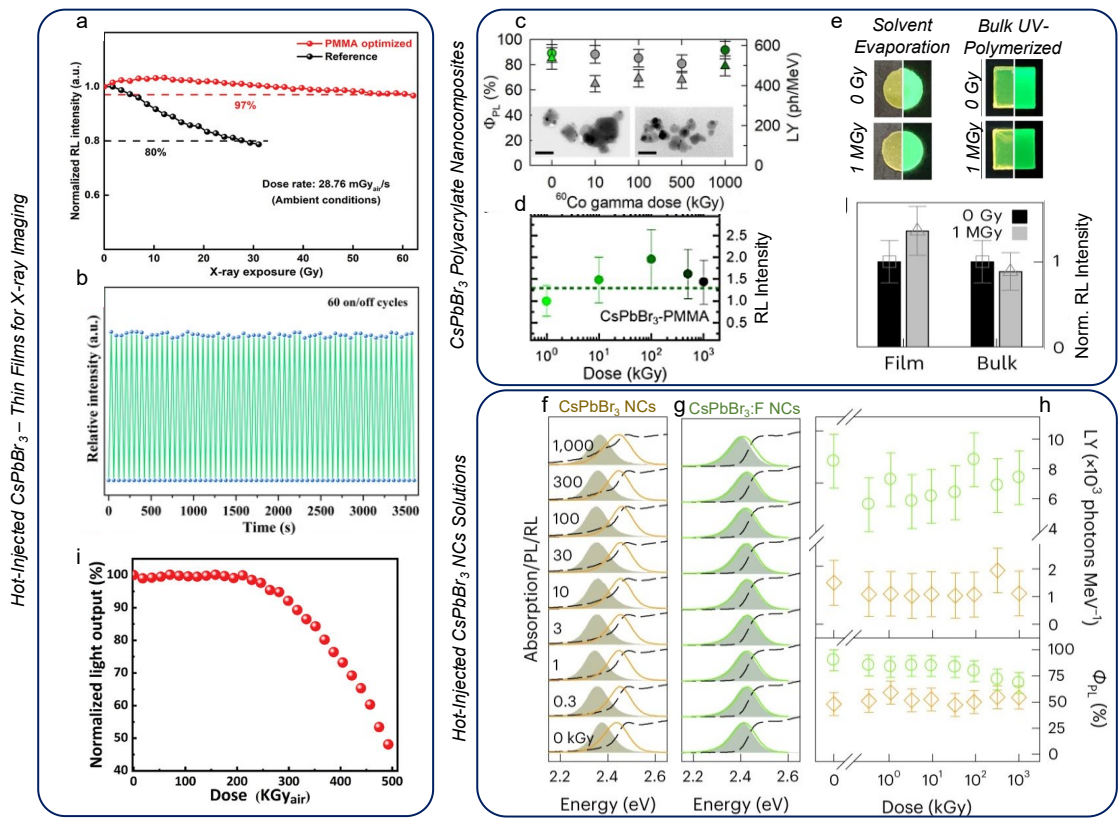

**Figure 9**: a) Normalized RL intensity of X-ray screen based on CsPbBr$_3$ PNCs monitored at increasing X-ray dose. Reproduced or adapted with permission from Ref ([226]). Copyright 2022, Wiley. b) RL stability of the CsPbBr$_3$-based scintillator under repeated cycles of 1 mGy·s$^{-1}$ X-ray excitation with a time interval of 30 s. Reproduced from Ref ([221]). Copyright 2020, American Chemistry Society. c) LY (circles) and corresponding Φ$_{PL}$ (triangles) as a function of cumulative dose for UV-polymerized CsPbBr$_3$ nanocomposite. Inset: TEM micrographs of 70 nm thin nanocomposite sections before and after irradiation (scale bar 50 nm). Reproduced from Ref ([63]). Copyright 2023, American Chemistry Society. d) RL intensity of LARP CsPbBr$_3$ PNCs and PMMA nanocomposite as a function of cumulative irradiation dose up to 1 MGy. The dashed line is a guide for the eyes. Reproduced rom Ref ([210]). Copyright 2024, American Chemistry Society. e) Photographs under ambient and UV light of PMMA/PLMA nanocomposites of CsPbBr$_3$ PNCs in form of thin film and bulk before and after 1 MGy of $^{60}$Co γ-dose. Below, normalized RL intensity of CsPbBr$_3$ PNCs showed in 'e' as a function of the delivered γ-ray dose. Reproduced or adapted with permission from Ref ([96]). Copyright 2022, Springer. Optical absorption (dashed lines), PL (solid lines) and RL (shaded areas) spectra of standard CsPbBr$_3$ (f) and CsPbBr$_3$:F PNCs (g) at increasing cumulative γ-ray doses from 0 Gy to 1 MGy (bottom to top). h) LY and respective relative RL intensity (top) and $\Phi_{PL}$ (bottom) of CsPbBr$_3$ (orange diamonds) and CsPbBr$_3$:F PNCs (green circles) as a function of cumulative dose. Panel f-h are reproduced or adapted with permission from Ref ([96]). Copyright 2022, Springer. i) RL intensity of CsPbBr$_3$ PNCs as a function of the $^{60}$Co γ-ray irradiation dose. Reproduced or adapted with permission from Ref ([225]). Copyright 2023, Wiley.

The origin of such resistance to radiation damage in PNCs is still debated, whereas in bulk perovskites self-healing processes have been observed under proton[92], soft X-rays[227], and γ-rays[228] irradiation showing that the ionic lattice mobility in LHP prompts the recycle of the degradation byproducts trapped in the crystal and results in a complete recovery of the optoelectronic properties. The impressive hardness of PNCs is of great interest in large experiments such as in Large Hadron Collider at CERN where the luminosity is already above $10^{34}$ cm$^{-1}$s$^{-1}$ and is expected to reach ~5·$10^{35}$ cm$^{-1}$s$^{-1}$ by 2035, resulting in an annual radiation load of over 1 MGy inside calorimeters and scintillating fiber tracking detectors[14, 229]. To date, organic plastic scintillators and optical fibers have been extensively used in high energy physics (HEP) experiments valued for their low cost and relatively fast detection characteristics, but suffer from prolonged radiation exposure resulting in light yield losses typically exceeding 50% above 100 kGy[14].

Along with all-inorganic lead halides PNCs, also Sb-based DP were demonstrated to have good radiation hardness, retaining their initial RL intensity upon exposed to continuous X-ray irradiation (dose rate of 0.5 Gy/s) up to almost 500 Gy of cumulated delivered dose[102]. A similar [SbCl$_6$] structure was also discussed by Zaffalon and co-worker[198] in a hybrid organic–inorganic zero-dimensional system, namely Gua$_3$SbCl$_6$, which consists of alternating layers of [SbCl$_6$]$^{3-}$ octahedra separated by N,N'-diphenylguanidinium (Gua$^+$, C$_{13}$H$_{14}$N$_3^+$) cations. These systems showed LY as high as 2300 ±500 ph·MeV$^{-1}$ (obtained by exciting the sample with a $^{57}$Co γ-ray source) which, combined with their excellent stability to prolonged exposure to ionizing radiation make Gua$_3$SbCl$_6$ PNCs promising candidates as active materials in high-energy detectors.

Timing Performance of PNC scintillators. A further key aspect of scintillation is timing, that is, the ability to produce a prompt pulse of light, fast enough to be used in applications that rely on the timing of events such as time-of-flight (TOF) based applications. These applications include HEP experiments, where high event rates in high-luminosity colliders such as High-Luminosity Large Hadron Collider (HL-LHC) and Future Circular Collider (FCC) require a fast response to mitigate the signal pile-up and localize the event vertexes, with double-pulse separations on the order of a few ns. Similarly, medical imaging with positron emission tomography (TOF-PET), where the coincident detection of two 511 keV photons emitted 180° back-to-back is used to reconstruct the position of positron annihilation events following the decay of a radiopharmaceutical tracer in the human body. In this sense, precise TOF information of the annihilation γ-photons dramatically improves the signal-to-noise ratio (SNR) in the reconstructed image by effectively rejecting background noise. As a result, precise timing of γ-photon arrival offers several performance advantages for TOF-PET scanners, including better image quality, shorter scan times, lower patient dose and higher spatial resolution[230].

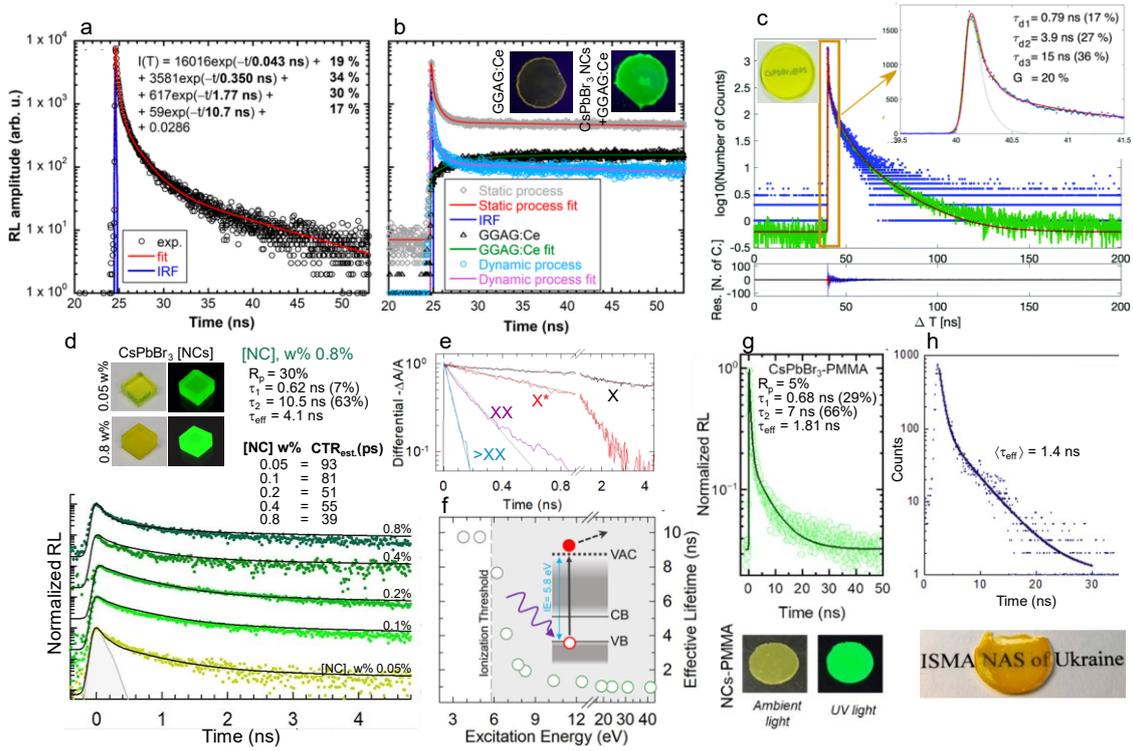

**Figure 10**: a) Radioluminescence decay of CsPbBr$_3$ PNCs casted onto silica glass together with the fitting parameters and the relative intensity of each decay component. b) Same as 'a' for CsPbBr$_3$ PNCs casted onto GGAG:Ce scintillating crystal. Inset: photographs of the two films under UV light illumination. Panels a and b are reproduced or adapted with permission from Ref ([157]). Copyright 2021, MPDI. c) RL decays of CsPbBr$_3$ PNCs embedded in a polystyrene matrix together with the detail of ultrafast dynamics and the extrapolated components. The G-labeled contribution correspond to an ultrafast decay component occurring within the instrumentation response time. Reproduced or adapted with permission from Ref ([155]). Copyright 2022, Royal Society of Chemistry. d) Radioluminescence decay traces of CsPbBr$_3$ PNCs nanocomposites (as shown in the photographs) at increasing NC loading. The R$_p$ contribution correspond to an ultrafast decay component similarly to the "G" value in panel 'c'. e) Differential TA curves of a representative nanocomposite from panel 'd' showing single-order, charged-order, bi-order, and higher order exciton dynamics. f) PL decay times for increasing excitation energy by synchrotron light up to E$_{EXC}$ = 19 eV. Inset: schematic depiction of the ionization of CsPbBr$_3$ PNCs upon excitation with the E$_{EXC}$ ≥ Ionization Energy. Panels d-f are reproduced from Ref ([63]). Copyright 2023, American Chemistry Society. g) RL decay trace for CsPbBr$_3$ PMMA nanocomposite (0.5 w%) excited with soft X-rays, together with the corresponding pictures under ambient and UV-light. The fitting curve is shown with solid line also reporting the fit parameters. Reproduced from Ref ([210]), Copyright 2024, American Chemistry Society. h) Scintillation decay curve of CsPbBr$_3$ PNCs composite film upon excitation by α-radiation from $^{239}$Pu. Reproduced or adapted with permission from Ref ([231]). Copyright 2023, Elsevier.

Specifically, compared to non-TOF techniques, the SNR improvement in TOF-PET scanners scales as[232]:

$$\frac{SNR_{TOF}}{SNR} \propto \sqrt{\frac{2D}{c \cdot CTR}}$$

where $D$ is the diameter of the object to be imaged, $c$ is the speed of light, and the coincidence time resolution (CTR) defines the temporal resolution of the system, expressed as the FWHM of the time spread distribution of the scintillating photons[233]. As a result, any sub-nanosecond CTR represents a significant advance over conventional non-TOF techniques, increasing image SNR by up to 3.5 times for the state-of-the-art commercial TOF-PET scanners. It follows that any appreciable progress in precision imaging is intimately linked to the improvement of this parameter, motivating the ambitious race towards 10 ps CTR[232]. Specifically, following the statistical description of the low-energy photons generated in the scintillator and their conversion into photoelectrons in the detector, the CTR can be modelled as follows[233]:

$$CTR \propto 3.33 \sqrt{\frac{\tau_r \tau_f}{\Phi_P}}$$

where $\tau_r$ and $\tau_f$ are the scintillator rise and fall luminescence time, respectively, and $\Phi_P$ is the total photoelectron yield per excitation energy which is proportional to the LY and accounts also for the geometric factors in detection assemblies (e.g. scintillation light outcoupling, waveguiding, surface back-reflections, and detector quantum efficiency). It follows that achieving ultimate time resolution with scintillator-based detectors requires a parallel effort on the light generation mechanisms, light transport optimization to reduce travel time jitter, the photoconversion system as well as the readout electronics[234].

> **The key to fabricating low CTR nanocrystal scintillators is to maximize ultrafast multi-exciton luminescence while minimizing reabsorption losses in high-density composites. Conditions that require conflicting strategies for which there is no definitive solution.**

In the HEP field, there is no comprehensive figure of merit for timing performance comparable to that of CTR. However, the timing resolution is determined by the variance with which enough photoelectrons are collected to provide a statistically significant signal. Despite this, HEP detectors share the fundamental principles of TOF-PET scintillators, aiming to combine ultra-fast scintillation with high light yield for higher timing resolution in radiation detection. To date, the lowest CTRs measured with coincidence setups under laboratory conditions at 511 keV are 127 ps for long crystals (3×3×20 mm$^3$) and 83 ps for small crystals (2×2×3 mm$^3$) of LYSO:Ce,Ca[235], while the best commercial TOF-PET scanners have a time resolution of 210 ps[236], corresponding to a spatial resolution of 30-40 mm along the line of response (LOR) before any image reconstruction algorithm is applied. As predicted by Lecoq *et al.*[237], the conventional scintillation mechanisms[238,239] and the resulting relaxation mechanisms fundamentally limit the achievable time resolution and require a paradigm shift in scintillating materials to pave the way towards 10 ps CTRs that would enable one millimeter diagnostic precision in TOF-PET. Among many concepts to enhance the production of ultrafast photons, such as Cherenkov emission[240], hot intraband luminescence[241], cross-luminescence[242], or probing the transient modulation of a material's refractive index[243], the quantum confinement effects in PNCs offer a promising combination of fast emission with relatively high light output[244,245]. The first reports on CsPbBr$_3$ PNCs as scintillation time-taggers came from Čuba and co-workers[155-157], aimed at extending the encouraging results from Turtos et al.[16] on heterostructures made of alternating thin layers of CdSe/CdS NPLs and polished LYSO:Ce single crystals. This configuration, known as *metascintillator* or *metapixel*, exploits the energy sharing between the two materials to add a relevant fraction of fast events on top of the intense and long-lived scintillation tail from the classic bulk scintillator resulting in a promising improvement of the time-tagging capabilities of the system[246]. For LHPs, this approach was first attempted by depositing a thin film of CsPbBr$_3$ PNCs on a scintillating wafer of LYSO:Ce[156] and GGAG:Ce[157], resulting in the characteristic RL decay profile shown in **Figure 10a**, **b**. More recently, a similar approach has been demonstrated by Pagano *et al*[247], who obtained a 10-fold improvement of the detector time resolution by casting μm-thick CsPbBr$_3$ PNCs films onto GAGG, BGO and LYSO crystals. Crucially, though, the observed luminescence enhancement of LHPs observed by Děcká *et al.* was largely a PL contribution following the optical excitation by the scintillation light of the underlying scintillator crystal instead of direct RL by the PNCs. This suggests that more complex approaches are required to achieve effective energy sharing of the electromagnetic shower, such as those recently proposed by Pagano *et al.*[248]. A more readily applicative analysis was carried out by the same group of Čuba[155] on highly concentrated nanocomposites - up to 10 wt% - of CsPbBr$_3$ PNC in polystyrene matrices. As shown in **Figure 10c**, when excited with X-rays at 10 keV, almost 40% of the total signal is due to sub-ns components, corresponding to a detector time resolution of 295 ps, which outperforms commercial LYSO:Ce crystals by a factor of two and is comparable to plastic EJ232, two of the best known commercially available scintillators. The origin of the sub-ns decay has been widely debated between the acceleration of the excitonic decay by nonradiative trapping or contributions by the recombination of multiexcitons formed under highly energetic excitation. The ambiguity has been recently lifted by Erroi *et al.*[63] who performed time-resolved RL and PL experiments using pulsed synchrotron light and ultrafast transient absorption (TA) measurements on CsPbBr$_3$ PNCs with PLQY>90% unambiguously ascribing the prompt RL emission to decay of double- and higher order multiexcitons. The authors used direct polymerization of polyacrylate matrixes to fabricate nanocomposites comprising increasing content of CsPbBr$_3$ PNCs and measured their *cw* and time resolved scintillation properties. Consistent with previous data by Čuba *et al.*, the authors observed ultrafast scintillation at any PNC loading, as highlighted in **Figure 10d** and measured the corresponding TA dynamics at increasing average exciton population (**Figure 10e**), which enabled them to isolate the kinetic contributions of excitonic species of increasing order (i.e., single excitons, charged excitons (trions), bi-excitons and higher order excitons) and to correlate them with the kinetic components observed in the RL decays. Through this combined spectroscopic and radiometric approach, they show that the ultrafast prompt RL component ($\tau_P < 200$ ps) - whose observation is limited by the time resolution of Time-correlated Single Photon Counting (TCSPC) setups - comprised biexciton and higher-order exciton species largely affected by non-radiative Auger processes. The intermediate dynamics ($\tau_{X^*} \sim 0.6$ ns) was instead found to originate from charged excitons following photoionization of the NCs inside the polymeric host, as demonstrated via time resolved PL experiments using synchrotron radiation with energy largely above the ionization energy of CsPbBr$_3$ PNCs (**Figure 10f**). This has profound implications for the general understanding of scintillation in LHP NCs, and suggested that suppression of Auger recombination affecting multiexciton decay would simultaneously increase the overall LY and proportionally slowdown the prompt RL dynamics, with the result that CTR timing performance is likely independent of the multiexciton yield and, consequently of the PNC size[52]. Rather, the key parameters for fabricating an ultra-low CTR nanoscintillator are the optical quality and the PNC loading of the nanocomposites to increase the average Z without introducing light scattering centres.

Consistently, the predicted CTR values in ref[63] (**Figure 10d**) improve with increasing PNC loading, decreasing from ~93 ps for the lowest PNC content (0.05 w%) to ~39 ps for the most concentrated nanocomposite (0.8 w%) featuring the highest LY value.

> **The key parameters for fabricating an ultra-low CTR nanoscintillator are the optical quality and the PNC loading of the nanocomposites to increase the average Z without introducing light scattering centers.**

Even more recently, Cova *et al.*[192] and Sorokin and co-workers[231] prepared respectively 0.5 w% and 2 w% loaded nanocomposite of CsPbBr$_3$ PNCs by solvent evaporation in PMMA. As shown in **Figure 10g, h**, they measured a largely ultrafast kinetics with effective scintillation time $\tau_{eff}$ ~1.81 ns by soft X-ray excitation and $\tau_{eff}$ ~1.4 ns, using $^{238}$Pu and $^{239}$Pu radionuclides as alpha particles excitation sources, in both cases ascribed to the recombination of multi- and charged-excitons formed under ionizing radiation, in full agreement with previous results[63, 155].

- **Summary and open challenges**

The study and application of PNCs as scintillating materials is a vibrant, yet still largely unexplored field, which so far showed great potential both in terms of the variety of materials that can be synthetized, for the multiplicity of emerging protocols for the fabrication of scintillating nanocomposites and in terms of their scalability to large volumes. Combined with their impressive radiation stability, ultrafast timing performance and promising scintillation efficiencies make this class of materials highly promising for scintillation application in many technological and scientific fields. Important challenges remain however open, such as the increase of the NCs loading to raise the density of the plastic matrix closer to inorganic crystals. This translates in the chemical challenge of finding effective surface chemistry strategies for compatibilizing PNCs with polymeric matrixes with minimal scattering losses and to stabilize them towards the high temperature protocols used in the living polymerization of high quality styrenes and acrylates and in extrusion processes. This is an open quest in the PNC field due to both the inherent instability of the weakly bound ligands and the ease of such ionic material to undergo phase transitions to non-emissive allotropes. Nonetheless, recent advancements in PNC encapsulation and resurfacing provide promising platform for overcoming this issue in the near future. On the fundamental photophysics side, the most challenging task is currently to maximize the outcoupling of the scintillation light by suppressing reabsorption losses in highly dense nanocomposites. Although this could be obtained with relatively slow scintillators based on PNCs with largely Stokes shifted, long-lived RL from trapped excitons or color centers, it is critical in fast scintillators that naturally require PNC with direct exciton decay featuring large oscillator strength, that unfortunately naturally corresponds to strong, energetically resonant, absorption. Addressing this aspect calls for innovative paradigms that conjugate fast recombination with sufficient decoupling between the luminescence and the dominant absorption.

**Acknowledgements**

This work was funded by Horizon Europe EIC Pathfinder program through project 101098649 – UNICORN and by the PRIN program of the Italian Ministry of University and Research (IRONSIDE project).

**Biographies**

**Abhinav Anand, Ph.D.** (*orcid.org/0000-0003-0530-5324*) received his Ph.D. in Materials Science and Nanotechnology from the University of Milano-Bicocca. He worked on colloidal synthesis and spectroscopic characterizations of nanocrystals exploring their applicative potential in ultrafast radiation detection in the same group before joining the esteemed Indian Institute of Science in Bangalore and finally moving to Vellore Institute of Technology as full-time Assistant Professor.

**Matteo L. Zaffalon, Ph.D.** (*orcid.org/0000-0002-1016-6413*) is a postdoctoral researcher at the University of Milano-Bicocca, where he earned his Ph.D. in 2022. His research activities focus on advanced spectroscopic studies of semiconducting colloidal nanocrystals for novel radiation detection schemes, including ultra-fast time-tagging and large-area devices.

**Andrea Erroi, Dr.** (*orcid.org/0000-0002-5128-7772*) is a PhD student in Materials Science and Nanotechnology at the University of Milano-Bicocca, where he received his master's degree in Materials Science in 2021. He works on the synthesis of nanocrystalline semiconductors and encapsulation in polymer matrices for scintillation applications.

**Francesca Cova, Ph.D.** (*orcid.org/0000-0001-7367-109X*) is assistant professor at the University of Milano – Bicocca, where she completed her Ph.D. in 2020 in Materials Science and Nanotechnology. Her research focuses on the scintillation properties of inorganic crystals, ceramics, colloidal nanostructures, and polymers, for their development and engineering in particle physics and medical diagnostics.

**Francesco Carulli, Ph.D.** (*orcid.org/0000-0002-8345-6606*) is an Assistant Professor in the Department of Materials Science at the University of Milan-Bicocca. He worked on fundamental and applied studies of colloidal nanocrystals as self-standing emitters and embedded into polymeric matrices. Recently, he has been investigating the scintillation properties of these materials for biological and high-energy physics applications.

**Sergio Brovelli, PhD, Professor** (*orcid.org/0000-0002-5993-855X*) received his Ph.D. from the University of Milano Bicocca (Italy) in 2006. He is now a Professor at the Department of Materials Science of the University of Milano Bicocca and co-founder and chair of the scientific committee of Glass to Power SpA. His research interests focus on the synthesis, manipulation, and spectroscopic investigation of solution-grown functional nanostructures and their application in photonic and optoelectronic devices.

References

(1) Bradley, W. G., History of medical imaging. *Proceedings of the American Philosophical Society* **2008,** *152* (3), 349-361.

(2) Ito, M.; Hong, S. J.; Lee, J. S., Positron emission tomography (PET) detectors with depth-of-interaction (DOI) capability. *Biomedical Engineering Letters* **2011,** *1* (2), 70-81.

(3) Crapanzano, R.; Secchi, V.; Villa, I., Co-Adjuvant Nanoparticles for Radiotherapy Treatments of Oncological Diseases. *Applied Science* **2021,** *11* (15), 7073.

(4) Hanke, R.; Fuchs, T.; Uhlmann, N., X-ray based methods for non-destructive testing and material characterization. *Nuclear Instruments and Methods in Physics Research Section A: Accelerators, Spectrometers, Detectors and Associated Equipment* **2008,** *591* (1), 14-18.

(5) Partridge, T.; Astolfo, A.; Shankar, S. S.; Vittoria, F. A.; Endrizzi, M.; Arridge, S.; Riley-Smith, T.; Haig, I. G.; Bate, D.; Olivo, A., Enhanced detection of threat materials by dark-field x-ray imaging combined with deep neural networks. *Nature Communications* **2022,** *13* (1), 4651.

(6) Faderl, N. In *Fast decay solid-state scintillators for high-speed x-ray imaging*, SPIE: 2019; pp 54-71.

(7) Kim, C.; Lee, W.; Melis, A.; Elmughrabi, A.; Lee, K.; Park, C.; Yeom, J.-Y. A Review of Inorganic Scintillation Crystals for Extreme Environments *Crystals* [Online], 2021, p. 669. https://www.mdpi.com/2073-4352/11/6/669.

(8) Fabjan, C. W.; Gianotti, F., Calorimetry for particle physics. *Reviews of Modern Physics* **2003,** *75* (4), 1243-1286.

(9) Knoll, G. F., *Radiation detection and measurement*. John Wiley & Sons: 2010.

(10) Birks, J. B., *The theory and practice of scintillation counting: International series of monographs in electronics and instrumentation*. Elsevier: 2013; Vol. 27.

(11) Wei, H.; DeSantis, D.; Wei, W.; Deng, Y.; Guo, D.; Savenije, T. J.; Cao, L.; Huang, J., Dopant compensation in alloyed $CH_3NH_3PbBr_{3-X}Cl_X$ perovskite single crystals for gamma-ray spectroscopy. *Nature Materials* **2017,** *16* (8), 826-833.

(12) Chen, Q.; Wu, J.; Ou, X.; Huang, B.; Almutlaq, J.; Zhumekenov, A. A.; Guan, X.; Han, S.; Liang, L.; Yi, Z.; Li, J.; Xie, X.; Wang, Y.; Li, Y.; Fan, D.; Teh, D. B. L.; All, A. H.; Mohammed, O. F.; Bakr, O. M.; Wu, T.; Bettinelli, M.; Yang, H.; Huang, W.; Liu, X., All-inorganic perovskite nanocrystal scintillators. *Nature* **2018,** *561* (7721), 88-93.

(13) Afanasiev, S. V.; Boyarintsev, A. Y.; Danilov, M. V.; Emeliantchik, I. F.; Ershov, Y. V.; Golutvin, I. A.; Grinyov, B. V.; Ibragimova, E.; Levchuk, L. G.; Litomin, A. V.; Makankin, A. M.; Malakhov, A. I.; Moisenz, P. V.; Nuritdinov, I.; Popov, V. F.; Rusinov, V. Y.; Shumeiko, N. M.; Smirnov, V. A.; Sorokin, P. V.; Tarkovskii, E. I.; Tashmetov, A.; Vasiliev, S. E.; Yuldashev, B.; Zamiatin, N. I.; Zhmurin, P. N., Light yield measurements of "finger" structured and unstructured scintillators after gamma and neutron irradiation. *Nuclear Instruments and Methods in Physics Research Section A: Accelerators, Spectrometers, Detectors and Associated Equipment* **2016,** *818*, 26-31.

(14) Kharzheev, Y. N., Radiation Hardness of Scintillation Detectors Based on Organic Plastic Scintillators and Optical Fibers. *Physics of Particles and Nuclei* **2019,** *50* (1), 42-76.

(15) Gandini, M.; Villa, I.; Beretta, M.; Gotti, C.; Imran, M.; Carulli, F.; Fantuzzi, E.; Sassi, M.; Zaffalon, M.; Brofferio, C.; Manna, L.; Beverina, L.; Vedda, A.; Fasoli, M.; Gironi, L.; Brovelli, S., Efficient, fast and reabsorption-free perovskite nanocrystal-based sensitized plastic scintillators. *Nature Nanotechnology* **2020,** *15* (6), 462-468.

(16) Turtos, R. M.; Gundacker, S.; Omelkov, S.; Mahler, B.; Khan, A. H.; Saaring, J.; Meng, Z.; Vasil'ev, A.; Dujardin, C.; Kirm, M.; Moreels, I.; Auffray, E.; Lecoq, P., On the use of CdSe scintillating nanoplatelets as time taggers for high-energy gamma detection. *npj 2D Materials and Applications* **2019,** *3* (1), 37.

(17) Turtos, R. M.; Gundacker, S.; Polovitsyn, A.; Christodoulou, S.; Salomoni, M.; Auffray, E.; Moreels, I.; Lecoq, P.; Grim, J. Q., Ultrafast emission from colloidal nanocrystals under pulsed X-ray excitation. *Journal of Instrumentation* **2016,** *11* (10), P10015.

(18) Cao, M.; Hu, C.; Wang, E., The first fluoride one-dimensional nanostructures: microemulsion-mediated hydrothermal synthesis of $BaF_2$ whiskers. *Journal of the American Chemical Society* **2003,** *125* (37), 11196-11197.

(19) Liu, C.; Hajagos, T. J.; Kishpaugh, D.; Jin, Y.; Hu, W.; Chen, Q.; Pei, Q., Facile Single-Precursor Synthesis and Surface Modification of Hafnium Oxide Nanoparticles for Nanocomposite γ-Ray Scintillators. *Advanced Functional Materials* **2015,** *25* (29), 4607-4616.

(20) Zhang, Y.-W.; Sun, X.; Si, R.; You, L.-P.; Yan, C.-H., Single-crystalline and monodisperse LaF3 triangular nanoplates from a single-source precursor. *Journal of the American Chemical Society* **2005,** *127* (10), 3260-3261.

(21) Sun, X.; Zhang, Y.-W.; Du, Y.-P.; Yan, Z.-G.; Si, R.; You, L.-P.; Yan, C.-H., From Trifluoroacetate Complex Precursors to Monodisperse Rare-Earth Fluoride and Oxyfluoride Nanocrystals with Diverse Shapes through Controlled Fluorination in Solution Phase. *Chemistry – A European Journal* **2007,** *13* (8), 2320-2332.

(22) Murray, C. B.; Norris, D. J.; Bawendi, M. G., Synthesis and characterization of nearly monodisperse CdE (E = sulfur, selenium, tellurium) semiconductor nanocrystallites. *Journal of the American Chemical Society* **1993,** *115* (19), 8706-8715.

(23) Peng, X.; Schlamp, M. C.; Kadavanich, A. V.; Alivisatos, A. P., Epitaxial Growth of Highly Luminescent CdSe/CdS Core/Shell Nanocrystals with Photostability and Electronic Accessibility. *Journal of the American Chemical Society* **1997,** *119* (30), 7019-7029.

(24) Wang, X.; Zhuang, J.; Peng, Q.; Li, Y., A general strategy for nanocrystal synthesis. *Nature* **2005,** *437* (7055), 121-124.

(25) Anand, A.; Zaffalon, M. L.; Cova, F.; Pinchetti, V.; Khan, A. H.; Carulli, F.; Brescia, R.; Meinardi, F.; Moreels, I.; Brovelli, S., Optical and Scintillation Properties of Record-Efficiency CdTe Nanoplatelets toward Radiation Detection Applications. *Nano Letters* **2022,** *22* (22), 8900-8907.

(26) Carulli, F.; Cova, F.; Gironi, L.; Meinardi, F.; Vedda, A.; Brovelli, S., Stokes Shift Engineered Mn:CdZnS/ZnS Nanocrystals as Reabsorption-Free Nanoscintillators in High Loading Polymer Composites. *Advanced Optical Materials* **2022,** *10* (13), 2200419.

(27) Gupta, S. K.; Mao, Y., Recent advances, challenges, and opportunities of inorganic nanoscintillators. *Frontiers of Optoelectronics* **2020,** *13* (2), 156-187.

(28) Heo, J. H.; Shin, D. H.; Park, J. K.; Kim, D. H.; Lee, S. J.; Im, S. H., High-Performance Next-Generation Perovskite Nanocrystal Scintillator for Nondestructive X-Ray Imaging. *Advanced Materials* **2018,** *30* (40), 1801743.

(29) Xu, Z.; Tang, X.; Liu, Y.; Zhang, Z.; Chen, W.; Liu, K.; Yuan, Z., $CsPbBr_3$ Quantum Dot Films with High Luminescence Efficiency and Irradiation Stability for Radioluminescent Nuclear Battery Application. *ACS Applied Materials & Interfaces* **2019,** *11* (15), 14191-14199.

(30) Moseley, O. D. I.; Doherty, T. A. S.; Parmee, R.; Anaya, M.; Stranks, S. D., Halide perovskites scintillators: unique promise and current limitations. *Journal of Materials Chemistry C* **2021,** *9* (35), 11588-11604.

(31) Wibowo, A.; Sheikh, M. A. K.; Diguna, L. J.; Ananda, M. B.; Marsudi, M. A.; Arramel, A.; Zeng, S.; Wong, L. J.; Birowosuto, M. D., Development and challenges in perovskite scintillators for high-resolution imaging and timing applications. *Communications Materials* **2023,** *4* (1), 21.

(32) Protesescu, L.; Yakunin, S.; Bodnarchuk, M. I.; Krieg, F.; Caputo, R.; Hendon, C. H.; Yang, R. X.; Walsh, A.; Kovalenko, M. V., Nanocrystals of Cesium Lead Halide Perovskites ($CsPbX_3$, X = Cl, Br, and I): Novel Optoelectronic Materials Showing Bright Emission with Wide Color Gamut. *Nano Letters* **2015,** *15* (6), 3692-3696.

(33) Yakunin, S.; Sytnyk, M.; Kriegner, D.; Shrestha, S.; Richter, M.; Matt, G. J.; Azimi, H.; Brabec, C. J.; Stangl, J.; Kovalenko, M. V.; Heiss, W., Detection of X-ray photons by solution-processed lead halide perovskites. *Nature Photonics* **2015,** *9* (7), 444-449.

(34) Wang, X.; Wu, Y.; Li, G.; Wu, J.; Zhang, X.; Li, Q.; Wang, B.; Chen, J.; Lei, W., Ultrafast Ionizing Radiation Detection by p–n Junctions Made with Single Crystals of Solution-Processed Perovskite. *Advanced Electronic Materials* **2018,** *4* (11), 1800237.

(35) Liu, J.; Shabbir, B.; Wang, C.; Wan, T.; Ou, Q.; Yu, P.; Tadich, A.; Jiao, X.; Chu, D.; Qi, D.; Li, D.; Kan, R.; Huang, Y.; Dong, Y.; Jasieniak, J.; Zhang, Y.; Bao, Q., Flexible, Printable Soft-X-Ray Detectors Based on All-Inorganic Perovskite Quantum Dots. *Advanced Materials* **2019,** *31* (30), 1901644.

(36) Kim, Y. C.; Kim, K. H.; Son, D.-Y.; Jeong, D.-N.; Seo, J.-Y.; Choi, Y. S.; Han, I. T.; Lee, S. Y.; Park, N.-G., Printable organometallic perovskite enables large-area, low-dose X-ray imaging. *Nature* **2017,** *550* (7674), 87-91.

(37) Yakunin, S.; Dirin, D. N.; Shynkarenko, Y.; Morad, V.; Cherniukh, I.; Nazarenko, O.; Kreil, D.; Nauser, T.; Kovalenko, M. V., Detection of gamma photons using solution-grown single crystals of hybrid lead halide perovskites. *Nature Photonics* **2016,** *10* (9), 585-589.

(38) Wei, H.; Fang, Y.; Mulligan, P.; Chuirazzi, W.; Fang, H.-H.; Wang, C.; Ecker, B. R.; Gao, Y.; Loi, M. A.; Cao, L.; Huang, J., Sensitive X-ray detectors made of methylammonium lead tribromide perovskite single crystals. *Nature Photonics* **2016,** *10* (5), 333-339.

(39) He, Y.; Matei, L.; Jung, H. J.; McCall, K. M.; Chen, M.; Stoumpos, C. C.; Liu, Z.; Peters, J. A.; Chung, D. Y.; Wessels, B. W.; Wasielewski, M. R.; Dravid, V. P.; Burger, A.; Kanatzidis, M. G., High spectral resolution of gamma-rays at room temperature by perovskite $CsPbBr_3$ single crystals. *Nature Communications* **2018,** *9* (1), 1609.

(40) Kakavelakis, G.; Gedda, M.; Panagiotopoulos, A.; Kymakis, E.; Anthopoulos, T. D.; Petridis, K., Metal Halide Perovskites for High-Energy Radiation Detection. *Advanced Science* **2020,** *7* (22), 2002098.

(41) Liu, F.; Wu, R.; Wei, J.; Nie, W.; Mohite, A. D.; Brovelli, S.; Manna, L.; Li, H., Recent Progress in Halide Perovskite Radiation Detectors for Gamma-Ray Spectroscopy. *ACS Energy Letters* **2022,** *7* (3), 1066-1085.

(42) Lim, J.; Hörantner, M. T.; Sakai, N.; Ball, J. M.; Mahesh, S.; Noel, N. K.; Lin, Y.-H.; Patel, J. B.; McMeekin, D. P.; Johnston, M. B.; Wenger, B.; Snaith, H. J., Elucidating the long-range charge carrier mobility in metal halide perovskite thin films. *Energy & Environmental Science* **2019,** *12* (1), 169-176.

(43) Lim, J.; Kober-Czerny, M.; Lin, Y.-H.; Ball, J. M.; Sakai, N.; Duijnstee, E. A.; Hong, M. J.; Labram, J. G.; Wenger, B.; Snaith, H. J., Long-range charge carrier mobility in metal halide perovskite thin-films and single crystals via transient photo-conductivity. *Nature Communications* **2022,** *13* (1), 4201.

(44) Gualdrón-Reyes, A. F.; Masi, S.; Mora-Seró, I., Progress in halide-perovskite nanocrystals with near-unity photoluminescence quantum yield. *Trends in Chemistry* **2021,** *3* (6), 499-511.

(45) Dey, A.; Ye, J.; De, A.; Debroye, E.; Ha, S. K.; Bladt, E.; Kshirsagar, A. S.; Wang, Z.; Yin, J.; Wang, Y.; Quan, L. N.; Yan, F.; Gao, M.; Li, X.; Shamsi, J.; Debnath, T.; Cao, M.; Scheel, M. A.; Kumar, S.; Steele, J. A.; Gerhard, M.; Chouhan, L.; Xu, K.; Wu, X.-g.; Li, Y.; Zhang, Y.; Dutta, A.; Han, C.; Vincon, I.; Rogach, A. L.; Nag, A.; Samanta, A.; Korgel, B. A.; Shih, C.-J.; Gamelin, D. R.; Son, D. H.; Zeng, H.; Zhong, H.; Sun, H.; Demir, H. V.; Scheblykin, I. G.; Mora-Seró, I.; Stolarczyk, J. K.; Zhang, J. Z.; Feldmann, J.; Hofkens, J.; Luther, J. M.; Pérez-Prieto, J.; Li, L.; Manna, L.; Bodnarchuk, M. I.; Kovalenko, M. V.; Roeffaers, M. B. J.; Pradhan, N.; Mohammed, O. F.; Bakr, O. M.; Yang, P.; Müller-Buschbaum, P.; Kamat, P. V.; Bao, Q.; Zhang, Q.; Krahne, R.; Galian, R. E.; Stranks, S. D.; Bals, S.; Biju, V.; Tisdale, W. A.; Yan, Y.; Hoye, R. L. Z.; Polavarapu, L., State of the Art and Prospects for Halide Perovskite Nanocrystals. *ACS Nano* **2021,** *15* (7), 10775–10981.

(46) Brown, A. A. M.; Damodaran, B.; Jiang, L.; Tey, J. N.; Pu, S. H.; Mathews, N.; Mhaisalkar, S. G., Lead Halide Perovskite Nanocrystals: Room Temperature Syntheses toward Commercial Viability. *Advanced Energy Materials* **2020,** *10* (34), 2001349.

(47) Akkerman, Q. A.; Gandini, M.; Di Stasio, F.; Rastogi, P.; Palazon, F.; Bertoni, G.; Ball, J. M.; Prato, M.; Petrozza, A.; Manna, L., Strongly emissive perovskite nanocrystal inks for high-voltage solar cells. *Nature Energy* **2016,** *2* (2), 16194.

(48) Zhang, F.; Chen, J.; Zhou, Y.; He, R.; Zheng, K., Effect of synthesis methods on photoluminescent properties for $CsPbBr_3$ nanocrystals: Hot injection method and conversion method. *Journal of Luminescence* **2020,** *220*, 117023.

(49) Maddalena, F.; Xie, A.; Chin, X. Y.; Begum, R.; Witkowski, M. E.; Makowski, M.; Mahler, B.; Drozdowski, W.; Springham, S. V.; Rawat, R. S.; Mathews, N.; Dujardin, C.; Birowosuto, M. D.; Dang, C., Deterministic Light Yield, Fast Scintillation, and Microcolumn Structures in Lead Halide Perovskite Nanocrystals. *The Journal of Physical Chemistry C* **2021,** *125* (25), 14082-14088.

(50) Clasen Hames, B.; Sánchez Sánchez, R.; Fakharuddin, A.; Mora-Seró, I., A Comparative Study of Light-Emitting Diodes Based on All-Inorganic Perovskite Nanoparticles ($CsPbBr_3$) Synthesized at Room Temperature and by a Hot-Injection Method. *ChemPlusChem* **2018,** *83* (4), 294-299.

(51) Akkerman, Q. A.; Nguyen, T. P. T.; Boehme, S. C.; Montanarella, F.; Dirin, D. N.; Wechsler, P.; Beiglböck, F.; Rainò, G.; Erni, R.; Katan, C.; Even, J.; Kovalenko, M. V., Controlling the nucleation and growth kinetics of lead halide perovskite quantum dots. *Science* **2022,** *377* (6613), 1406-1412.

(52) Shamsi, J.; Urban, A. S.; Imran, M.; De Trizio, L.; Manna, L., Metal Halide Perovskite Nanocrystals: Synthesis, Post-Synthesis Modifications, and Their Optical Properties. *Chemical Reviews* **2019,** *119* (5), 3296-3348.

(53) Smock, S. R.; Chen, Y.; Rossini, A. J.; Brutchey, R. L., The Surface Chemistry and Structure of Colloidal Lead Halide Perovskite Nanocrystals. *Accounts of Chemical Research* **2021,** *54* (3), 707-718.

(54) De Trizio, L.; Infante, I.; Manna, L., Surface Chemistry of Lead Halide Perovskite Colloidal Nanocrystals. *Accounts of Chemical Research* **2023,** *56* (13), 1815-1825.

(55) Yang, D.; Li, X.; Zeng, H., Surface Chemistry of All Inorganic Halide Perovskite Nanocrystals: Passivation Mechanism and Stability. *Advanced Materials Interfaces* **2018,** *5* (8), 1701662.

(56) Chen, W.; Liu, Y.; Yuan, Z.; Xu, Z.; Zhang, Z.; Liu, K.; Jin, Z.; Tang, X., X-ray radioluminescence effect of all-inorganic halide perovskite $CsPbBr_3$ quantum dots. *Journal of Radioanalytical and Nuclear Chemistry* **2017,** *314* (3), 2327-2337.

(57) Zhang, Y.; Sun, R.; Ou, X.; Fu, K.; Chen, Q.; Ding, Y.; Xu, L.-J.; Liu, L.; Han, Y.; Malko, A. V.; Liu, X.; Yang, H.; Bakr, O. M.; Liu, H.; Mohammed, O. F., Metal Halide Perovskite Nanosheet for X-ray High-Resolution Scintillation Imaging Screens. *ACS Nano* **2019,** *13* (2), 2520-2525.

(58) Wang, L.; Fu, K.; Sun, R.; Lian, H.; Hu, X.; Zhang, Y., Ultra-stable $CsPbBr_3$ Perovskite Nanosheets for X-Ray Imaging Screen. *Nano-Micro Letters* **2019,** *11* (1), 52.

(59) Montanarella, F.; McCall, K. M.; Sakhatskyi, K.; Yakunin, S.; Trtik, P.; Bernasconi, C.; Cherniukh, I.; Mannes, D.; Bodnarchuk, M. I.; Strobl, M.; Walfort, B.; Kovalenko, M. V., Highly Concentrated, Zwitterionic Ligand-Capped $Mn^{2+}:CsPb(Br_XCl_{1-X})_3$ Nanocrystals as Bright Scintillators for Fast Neutron Imaging. *ACS Energy Letters* **2021,** *6* (12), 4365-4373.

(60) Lecoq, P., Development of new scintillators for medical applications. *Nuclear Instruments and Methods in Physics Research Section A: Accelerators, Spectrometers, Detectors and Associated Equipment* **2016,** *809*, 130-139.

(61) Wei, W.; Zhang, Y.; Xu, Q.; Wei, H.; Fang, Y.; Wang, Q.; Deng, Y.; Li, T.; Gruverman, A.; Cao, L., Monolithic integration of hybrid perovskite single crystals with heterogenous substrate for highly sensitive X-ray imaging. *Nature Photonics* **2017,** *11* (5), 315-321.

(62) Glodo, J.; Wang, Y.; Shawgo, R.; Brecher, C.; Hawrami, R. H.; Tower, J.; Shah, K. S., New developments in scintillators for security applications. *Physics Procedia* **2017,** *90*, 285-290.

(63) Erroi, A.; Mecca, S.; Zaffalon, M. L.; Frank, I.; Carulli, F.; Cemmi, A.; Di Sarcina, I.; Debellis, D.; Rossi, F.; Cova, F.; Pauwels, K.; Mauri, M.; Perego, J.; Pinchetti, V.; Comotti, A.; Meinardi, F.; Vedda, A.; Auffray, E.; Beverina, L.; Brovelli, S., Ultrafast and Radiation-Hard Lead Halide Perovskite Nanocomposite Scintillators. *ACS Energy Letters* **2023,** *8* (9), 3883-3894.

(64) Coropceanu, I.; Bawendi, M. G., Core/shell quantum dot based luminescent solar concentrators with reduced reabsorption and enhanced efficiency. *Nano Letters* **2014,** *14* (7), 4097-4101.

(65) Meinardi, F.; Colombo, A.; Velizhanin, K. A.; Simonutti, R.; Lorenzon, M.; Beverina, L.; Viswanatha, R.; Klimov, V. I.; Brovelli, S., Large-area luminescent solar concentrators based on 'Stokes-shift-engineered' nanocrystals in a mass-polymerized PMMA matrix. *Nature Photonics* **2014,** *8* (5), 392-399.

(66) Meinardi, F.; McDaniel, H.; Carulli, F.; Colombo, A.; Velizhanin, K. A.; Makarov, N. S.; Simonutti, R.; Klimov, V. I.; Brovelli, S., Highly efficient large-area colourless luminescent solar concentrators using heavy-metal-free colloidal quantum dots. *Nature Nanotechnology* **2015,** *10* (10), 878-885.

(67) Zhou, Y.; Wang, X.; He, T.; Yang, H.; Yang, C.; Shao, B.; Gutiérrez-Arzaluz, L.; Bakr, O. M.; Zhang, Y.; Mohammed, O. F., Large-Area Perovskite-Related Copper Halide Film for High-Resolution Flexible X-ray Imaging Scintillation Screens. *ACS Energy Letters* **2022,** *7* (2), 844-846.

(68) Mir, W. J.; Mahor, Y.; Lohar, A.; Jagadeeswararao, M.; Das, S.; Mahamuni, S.; Nag, A., Postsynthesis doping of Mn and Yb into $CsPbX_3$ (X= Cl, Br, or I) perovskite nanocrystals for downconversion emission. *Chemistry of Materials* **2018,** *30* (22), 8170-8178.

(69) Meinardi, F.; Akkerman, Q. A.; Bruni, F.; Park, S.; Mauri, M.; Dang, Z.; Manna, L.; Brovelli, S., Doped halide perovskite nanocrystals for reabsorption-free luminescent solar concentrators. *ACS energy letters* **2017,** *2* (10), 2368-2377.

(70) Liu, W.; Lin, Q.; Li, H.; Wu, K.; Robel, I.; Pietryga, J. M.; Klimov, V. I., $Mn^{2+}$-doped lead halide perovskite nanocrystals with dual-color emission controlled by halide content. *Journal of the American Chemical Society* **2016,** *138* (45), 14954-14961.

(71) Das Adhikari, S.; Guria, A. K.; Pradhan, N., Insights of doping and the photoluminescence properties of Mn-doped perovskite nanocrystals. *The Journal of Physical Chemistry Letters* **2019,** *10* (9), 2250-2257.

(72) Guria, A. K.; Dutta, S. K.; Adhikari, S. D.; Pradhan, N., Doping $Mn^{2+}$ in lead halide perovskite nanocrystals: successes and challenges. *ACS Energy Letters* **2017,** *2* (5), 1014-1021.

(73) Huang, G.; Wang, C.; Xu, S.; Zong, S.; Lu, J.; Wang, Z.; Lu, C.; Cui, Y., Postsynthetic Doping of $MnCl_2$ Molecules into Preformed $CsPbBr_3$ Perovskite Nanocrystals via a Halide Exchange-Driven Cation Exchange. *Advanced materials* **2017,** *29* (29), 1700095.

(74) Mir, W. J.; Jagadeeswararao, M.; Das, S.; Nag, A., Colloidal Mn-doped cesium lead halide perovskite nanoplatelets. *ACS Energy Letters* **2017,** *2* (3), 537-543.

(75) Ji, S.; Yuan, X.; Cao, S.; Ji, W.; Zhang, H.; Wang, Y.; Li, H.; Zhao, J.; Zou, B., Near-unity red $Mn^{2+}$ photoluminescence quantum yield of doped $CsPbCl_3$ nanocrystals with Cd incorporation. *The Journal of Physical Chemistry Letters* **2020,** *11* (6), 2142-2149.

(76) Das Adhikari, S.; Behera, R. K.; Bera, S.; Pradhan, N., Presence of metal chloride for minimizing the halide deficiency and maximizing the doping efficiency in Mn (II)-doped $CsPbCl_3$ nanocrystals. *The Journal of Physical Chemistry Letters* **2019,** *10* (7), 1530-1536.

(77) Paul, S.; Bladt, E.; Richter, A. F.; Döblinger, M.; Tong, Y.; Huang, H.; Dey, A.; Bals, S.; Debnath, T.; Polavarapu, L., Manganese-Doping-Induced Quantum Confinement within Host Perovskite Nanocrystals through Ruddlesden–Popper Defects. *Angewandte Chemie International Edition* **2020,** *59* (17), 6794-6799.

(78) Dagnall, K. A.; Conley, A. M.; Yoon, L. U.; Rajeev, H. S.; Lee, S.-H.; Choi, J. J., Ytterbium-Doped Cesium Lead Chloride Perovskite as an X-ray Scintillator with High Light Yield. *ACS Omega* **2022,** *7* (24), 20968-20974.

(79) Zhou, D.; Liu, D.; Pan, G.; Chen, X.; Li, D.; Xu, W.; Bai, X.; Song, H., Cerium and Ytterbium Codoped Halide Perovskite Quantum Dots: A Novel and Efficient Downconverter for Improving the Performance of Silicon Solar Cells. *Advanced Materials* **2017,** *29* (42), 1704149.

(80) Krieg, F.; Ong, Q. K.; Burian, M.; Rainò, G.; Naumenko, D.; Amenitsch, H.; Süess, A.; Grotevent, M. J.; Krumeich, F.; Bodnarchuk, M. I., Stable ultraconcentrated and ultradilute colloids of CsPbX$_3$ (X= Cl, Br) nanocrystals using natural lecithin as a capping ligand. *Journal of the American Chemical Society* **2019,** *141* (50), 19839-19849.

(81) McCall, K. M.; Sakhatskyi, K.; Lehmann, E.; Walfort, B.; Losko, A. S.; Montanarella, F.; Bodnarchuk, M. I.; Krieg, F.; Kelestemur, Y.; Mannes, D.; Shynkarenko, Y.; Yakunin, S.; Kovalenko, M. V., Fast Neutron Imaging with Semiconductor Nanocrystal Scintillators. *ACS Nano* **2020,** *14* (11), 14686-14697.

(82) Cao, F.; Yu, D.; Ma, W.; Xu, X.; Cai, B.; Yang, Y. M.; Liu, S.; He, L.; Ke, Y.; Lan, S.; Choy, K.-L.; Zeng, H., Shining Emitter in a Stable Host: Design of Halide Perovskite Scintillators for X-ray Imaging from Commercial Concept. *ACS Nano* **2020,** *14* (5), 5183-5193.

(83) Xu, Q.; Wang, J.; Shao, W.; Ouyang, X.; Wang, X.; Zhang, X.; Guo, Y.; Ouyang, X., A solution-processed zero-dimensional all-inorganic perovskite scintillator for high resolution gamma-ray spectroscopy detection. *Nanoscale* **2020,** *12* (17), 9727-9732.

(84) Ren, C.; Li, Z.; Huang, L.; Xiong, X.; Nie, Z.; Yang, Y.; Zhu, W.; Yang, W.; Wang, L., Confinement of all-inorganic perovskite quantum dots assembled in metal–organic frameworks for ultrafast scintillator application. *Nanoscale* **2022,** *14* (11), 4216-4224.

(85) Zhang, H.; Yang, Z.; Zhou, M.; Zhao, L.; Jiang, T.; Yang, H.; Yu, X.; Qiu, J.; Yang, Y.; Xu, X., Reproducible X-ray Imaging with a Perovskite Nanocrystal Scintillator Embedded in a Transparent Amorphous Network Structure. *Advanced Materials* **2021,** *33* (40), 2102529.

(86) Ryu, I.; Ryu, J.-Y.; Choe, G.; Kwon, H.; Park, H.; Cho, Y.-S.; Du, R.; Yim, S., In Vivo Plain X-Ray Imaging of Cancer Using Perovskite Quantum Dot Scintillators. *Advanced Functional Materials* **2021,** *31* (34), 2102334.

(87) Zhang, Q.; Zheng, W.; Wan, Q.; Liu, M.; Feng, X.; Kong, L.; Li, L., Confined Synthesis of Stable and Uniform CsPbBr$_3$ Nanocrystals with High Quantum Yield up to 90% by High Temperature Solid-State Reaction. **2021,** *9* (11), 2002130.

(88) Gómez Andrade, V. A.; Herrera Martínez, W. O.; Redondo, F.; Correa Guerrero, N. B.; Roncaroli, F.; Perez, M. D., Fe and Ti metal-organic frameworks: Towards tailored materials for photovoltaic applications. *Applied Materials Today* **2021,** *22*, 100915.

(89) Kim, H.; Rao, S. R.; Kapustin, E. A.; Zhao, L.; Yang, S.; Yaghi, O. M.; Wang, E. N., Adsorption-based atmospheric water harvesting device for arid climates. *Nature Communications* **2018,** *9* (1), 1191.

(90) Hou, J.; Wang, Z.; Chen, P.; Chen, V.; Cheetham, A. K.; Wang, L., Intermarriage of Halide Perovskites and Metal-Organic Framework Crystals. *Angewandte Chemie International Edition* **2020,** *59* (44), 19434-19449.

(91) Li, H.; Yang, H.; Yuan, R.; Sun, Z.; Yang, Y.; Zhao, J.; Li, Q.; Zhang, Z., Ultrahigh Spatial Resolution, Fast Decay, and Stable X-Ray Scintillation Screen through Assembling CsPbBr3 Nanocrystals Arrays in Anodized Aluminum Oxide. *Advanced Optical Materials* **2021,** *9* (24), 2101297.

(92) Yu, Y.; Zhang, F.; Yu, H., Self-healing perovskite solar cells. *Solar Energy* **2020,** *209*, 408-414.

(93) He, M.; Zhang, Q.; Carulli, F.; Erroi, A.; Wei, W.; Kong, L.; Yuan, C.; Wan, Q.; Liu, M.; Liao, X.; Zhan, W.; Han, L.; Guo, X.; Brovelli, S.; Li, L., Ultra-stable, Solution-Processable CsPbBr$_3$-

SiO$_2$ Nanospheres for Highly Efficient Color Conversion in Micro Light-Emitting Diodes. *ACS Energy Letters* **2023,** *8* (1), 151-158.

(94) Carulli, F.; He, M.; Cova, F.; Erroi, A.; Li, L.; Brovelli, S., Silica-Encapsulated Perovskite Nanocrystals for X-ray-Activated Singlet Oxygen Production and Radiotherapy Application. *ACS Energy Letters* **2023,** *8* (4), 1795-1802.

(95) Yang, D.; Xu, Z.; Gong, C.; Su, L.; Li, X.; Tang, X.; Geng, D.; Meng, C.; Xu, F.; Zeng, H., Armor-like passivated CsPbBr$_3$ quantum dots: boosted stability with hand-in-hand ligands and enhanced performance of nuclear batteries. *Journal of Materials Chemistry A* **2021,** *9* (13), 8772-8781.

(96) Zaffalon, M. L.; Cova, F.; Liu, M.; Cemmi, A.; Di Sarcina, I.; Rossi, F.; Carulli, F.; Erroi, A.; Rodà, C.; Perego, J.; Comotti, A.; Fasoli, M.; Meinardi, F.; Li, L.; Vedda, A.; Brovelli, S., Extreme γ-ray radiation hardness and high scintillation yield in perovskite nanocrystals. *Nature Photonics* **2022,** *16* (12), 860-868.

(97) Liu, M.; Wan, Q.; Wang, H.; Carulli, F.; Sun, X.; Zheng, W.; Kong, L.; Zhang, Q.; Zhang, C.; Zhang, Q.; Brovelli, S.; Li, L., Suppression of temperature quenching in perovskite nanocrystals for efficient and thermally stable light-emitting diodes. *Nature Photonics* **2021,** *15* (5), 379-385.

(98) Wang, S.; Huang, W.; Liu, X.; Wang, S.; Ye, H.; Yu, S.; Song, X.; Liu, Z.; Wang, P.; Yang, T.; Chu, D.; Gou, J.; Yuan, M.; Chen, L.; Su, B.; Liu, S.; Zhao, K., Ruddlesden–Popper Perovskite Nanocrystals Stabilized in Mesoporous Silica with Efficient Carrier Dynamics for Flexible X-Ray Scintillator. *Advanced Functional Materials* **2023,** *33* (3), 2210765.

(99) Xia, K.; Ran, P.; Wang, W.; Yu, J.; Xu, G.; Wang, K.; Pi, X.; He, Q.; Yang, Y.; Pan, J., In Situ Preparation of High-Quality Flexible Manganese-Halide Scintillator Films for X-Ray Imaging. *Advanced Optical Materials* **2022,** *10* (20), 2201028.

(100) Zhou, J.; An, K.; He, P.; Yang, J.; Zhou, C.; Luo, Y.; Kang, W.; Hu, W.; Feng, P.; Zhou, M.; Tang, X., Solution-Processed Lead-Free Perovskite Nanocrystal Scintillators for High-Resolution X-Ray CT Imaging. *Advanced Optical Materials* **2021,** *9* (11), 2002144.

(101) Lian, L.; Zheng, M.; Zhang, W.; Yin, L.; Du, X.; Zhang, P.; Zhang, X.; Gao, J.; Zhang, D.; Gao, L.; Niu, G.; Song, H.; Chen, R.; Lan, X.; Tang, J.; Zhang, J., Efficient and Reabsorption-Free Radioluminescence in Cs$_3$Cu$_2$I$_5$ Nanocrystals with Self-Trapped Excitons. *Advanced Science* **2020,** *7* (11), 2000195.

(102) Zhu, D.; Zaffalon, M. L.; Zito, J.; Cova, F.; Meinardi, F.; De Trizio, L.; Infante, I.; Brovelli, S.; Manna, L., Sb-Doped Metal Halide Nanocrystals: A 0D versus 3D Comparison. *ACS Energy Letters* **2021,** *6* (6), 2283-2292.

(103) Smith, M. D.; Connor, B. A.; Karunadasa, H. I., Tuning the luminescence of layered halide perovskites. *Chemical Reviews* **2019,** *119* (5), 3104-3139.

(104) Yuan, M.; Quan, L. N.; Comin, R.; Walters, G.; Sabatini, R.; Voznyy, O.; Hoogland, S.; Zhao, Y.; Beauregard, E. M.; Kanjanaboos, P., Perovskite energy funnels for efficient light-emitting diodes. *Nature Nanotechnology* **2016,** *11* (10), 872-877.

(105) Byun, J.; Cho, H.; Wolf, C.; Jang, M.; Sadhanala, A.; Friend, R. H.; Yang, H.; Lee, T.-W., Perovskite Light-Emitting Diodes: Efficient Visible Quasi-2D Perovskite Light-Emitting Diodes. *Advanced Materials* **2016,** *28* (34), 7550-7550.

(106) Giovanni, D.; Lim, J. W. M.; Yuan, Z.; Lim, S. S.; Righetto, M.; Qing, J.; Zhang, Q.; Dewi, H. A.; Gao, F.; Mhaisalkar, S. G.; Mathews, N.; Sum, T. C., Ultrafast long-range spin-funneling in solution-processed Ruddlesden–Popper halide perovskites. *Nature Communications* **2019,** *10* (1), 3456.

(107) Papavassiliou, G. C., Three-and low-dimensional inorganic semiconductors. *Progress in Solid State Chemistry* **1997,** *25* (3-4), 125-270.

(108) Shinozaki, K.; Kawano, N., Rapid Synthesis of Quantum-Sized Organic–Inorganic Perovskite Nanocrystals in Glass. *Scientific Reports* **2020,** *10* (1), 1237.

(109) Leijtens, T.; Eperon, G. E.; Noel, N. K.; Habisreutinger, S. N.; Petrozza, A.; Snaith, H. J., Stability of metal halide perovskite solar cells. *Advanced Energy Materials* **2015,** *5* (20), 1500963.

(110) Zhang, H.; Wang, X.; Liao, Q.; Xu, Z.; Li, H.; Zheng, L.; Fu, H., Embedding perovskite nanocrystals into a polymer matrix for tunable luminescence probes in cell imaging. *Advanced Functional Materials* **2017,** *27* (7), 1604382.

(111) Huang, H.; Chen, B.; Wang, Z.; Hung, T. F.; Susha, A. S.; Zhong, H.; Rogach, A. L., Water resistant $CsPbX_3$ nanocrystals coated with polyhedral oligomeric silsesquioxane and their use as solid state luminophores in all-perovskite white light-emitting devices. *Chemical Science* **2016,** *7* (9), 5699-5703.

(112) Sun, J. Y.; Rabouw, F. T.; Yang, X. F.; Huang, X. Y.; Jing, X. P.; Ye, S.; Zhang, Q. Y., Facile two-step synthesis of all-inorganic perovskite $CsPbX_3$ (X= Cl, Br, and I) zeolite-Y composite phosphors for potential backlight display application. *Advanced Functional Materials* **2017,** *27* (45), 1704371.

(113) Dang, Z.; Shamsi, J.; Palazon, F.; Imran, M.; Akkerman, Q. A.; Park, S.; Bertoni, G.; Prato, M.; Brescia, R.; Manna, L., In situ transmission electron microscopy study of electron beam-induced transformations in colloidal cesium lead halide perovskite nanocrystals. *ACS Nano* **2017,** *11* (2), 2124-2132.

(114) Parobek, D.; Dong, Y.; Qiao, T.; Rossi, D.; Son, D. H., Photoinduced anion exchange in cesium lead halide perovskite nanocrystals. *Journal of the American Chemical Society* **2017,** *139* (12), 4358-4361.

(115) Kim, Y.; Yassitepe, E.; Voznyy, O.; Comin, R.; Walters, G.; Gong, X.; Kanjanaboos, P.; Nogueira, A. F.; Sargent, E. H., Efficient luminescence from perovskite quantum dot solids. *ACS Applied Materials & Interfaces* **2015,** *7* (45), 25007-25013.

(116) Loiudice, A.; Saris, S.; Oveisi, E.; Alexander, D. T. L.; Buonsanti, R., $CsPbBr_3$ QD/$AlO_X$ inorganic nanocomposites with exceptional stability in water, light, and heat. *Angewandte Chemie International Edition* **2017,** *56* (36), 10696-10701.

(117) Polavarapu, L.; Nickel, B.; Feldmann, J.; Urban, A. S., Advances in Quantum-Confined Perovskite Nanocrystals for Optoelectronics. *Advanced Energy Materials* **2017,** *7* (16), 1700267.

(118) Balazs, A. C.; Emrick, T.; Russell, T. P., Nanoparticle Polymer Composites: Where Two Small Worlds Meet. *Science* **2006,** *314* (5802), 1107-1110.

(119) Lü, C.; Yang, B., High refractive index organic–inorganic nanocomposites: design, synthesis and application. *Journal of Materials Chemistry* **2009,** *19* (19), 2884-2901.

(120) Novak, B. M., Hybrid Nanocomposite Materials—between inorganic glasses and organic polymers. *Advanced Materials* **1993,** *5* (6), 422-433.

(121) Mutin, P. H.; Guerrero, G.; Vioux, A., Hybrid materials from organophosphorus coupling molecules. *Journal of Materials Chemistry* **2005,** *15* (35-36), 3761-3768.

(122) Boles, M. A.; Ling, D.; Hyeon, T.; Talapin, D. V., The surface science of nanocrystals. *Nature Materials* **2016,** *15* (2), 141-153.

(123) Li, Y.; Krentz, T. M.; Wang, L.; Benicewicz, B. C.; Schadler, L. S., Ligand Engineering of Polymer Nanocomposites: From the Simple to the Complex. *ACS Applied Materials & Interfaces* **2014,** *6* (9), 6005-6021.

(124) Kim, H.; Hight-Huf, N.; Kang, J.-H.; Bisnoff, P.; Sundararajan, S.; Thompson, T.; Barnes, M.; Hayward, R. C.; Emrick, T., Polymer Zwitterions for Stabilization of $CsPbBr_3$ Perovskite Nanoparticles and Nanocomposite Films. *Angewandte Chemie International Edition* **2020,** *59* (27), 10802-10806.

(125) Pan, A.; Wang, J.; Jurow, M. J.; Jia, M.; Liu, Y.; Wu, Y.; Zhang, Y.; He, L.; Liu, Y., General Strategy for the Preparation of Stable Luminous Nanocomposite Inks Using Chemically Addressable $CsPbX_3$ Peroskite Nanocrystals. *Chemistry of Materials* **2018,** *30* (8), 2771-2780.

(126) Hou, S.; Guo, Y.; Tang, Y.; Quan, Q., Synthesis and Stabilization of Colloidal Perovskite Nanocrystals by Multidentate Polymer Micelles. *ACS Applied Materials & Interfaces* **2017,** *9* (22), 18417-18422.

(127) Tong, J.; Wu, J.; Shen, W.; Zhang, Y.; Liu, Y.; Zhang, T.; Nie, S.; Deng, Z., Direct Hot-Injection Synthesis of Lead Halide Perovskite Nanocubes in Acrylic Monomers for Ultrastable and Bright Nanocrystal–Polymer Composite Films. *ACS Applied Materials & Interfaces* **2019,** *11* (9), 9317-9325.

(128) He, J.; He, Z.; Towers, A.; Zhan, T.; Chen, H.; Zhou, L.; Zhang, C.; Chen, R.; Sun, T.; Gesquiere, A. J.; Wu, S.-T.; Dong, Y., Ligand assisted swelling–deswelling microencapsulation (LASDM) for stable, color tunable perovskite–polymer composites. *Nanoscale Advances* **2020,** *2* (5), 2034-2043.

(129) Zhao, C.; Song, H.; Chen, Y.; Xiong, W.; Hu, M.; Wu, Y.; Zhang, Y.; He, L.; Liu, Y.; Pan, A., Stable and Recyclable Photocatalysts of $CsPbBr_3$@MSNs Nanocomposites for Photoinduced Electron Transfer RAFT Polymerization. *ACS Energy Letters* **2022,** *7* (12), 4389-4397.

(130) Peng, J.; Khan, J. I.; Liu, W.; Ugur, E.; Duong, T.; Wu, Y.; Shen, H.; Wang, K.; Dang, H.; Aydin, E.; Yang, X.; Wan, Y.; Weber, K. J.; Catchpole, K. R.; Laquai, F.; De Wolf, S.; White, T. P., A Universal Double-Side Passivation for High Open-Circuit Voltage in Perovskite Solar Cells: Role of Carbonyl Groups in Poly(methyl methacrylate). *Advanced Energy Materials* **2018,** *8* (30), 1801208.

(131) Sun, H.; Yang, Z.; Wei, M.; Sun, W.; Li, X.; Ye, S.; Zhao, Y.; Tan, H.; Kynaston, E. L.; Schon, T. B.; Yan, H.; Lu, Z.-H.; Ozin, G. A.; Sargent, E. H.; Seferos, D. S., Chemically Addressable Perovskite Nanocrystals for Light-Emitting Applications. *Advanced Materials* **2017,** *29* (34), 1701153.

(132) Kuo, S.-W.; Chang, F.-C., POSS related polymer nanocomposites. *Progress in Polymer Science* **2011,** *36* (12), 1649-1696.

(133) Ramirez, S. M.; Diaz, Y. J.; Sahagun, C. M.; Duff, M. W.; Lawal, O. B.; Iacono, S. T.; Mabry, J. M., Reversible addition–fragmentation chain transfer (RAFT) copolymerization of fluoroalkyl polyhedral oligomeric silsesquioxane (F-POSS) macromers. *Polymer Chemistry* **2013,** *4* (7), 2230-2234.

(134) Xu, N.; Stark, E. J.; Dvornic, P. R.; Meier, D. J.; Hu, J.; Hartmann-Thompson, C., Hyperbranched Polycarbosiloxanes and Polysiloxanes with Octafunctional Polyhedral Oligomeric Silsesquioxane (POSS) Branch Points. *Macromolecules* **2012,** *45* (11), 4730-4739.

(135) Cordes, D. B.; Lickiss, P. D.; Rataboul, F., Recent Developments in the Chemistry of Cubic Polyhedral Oligosilsesquioxanes. *Chemical Reviews* **2010,** *110* (4), 2081-2173.

(136) Wang, F.; Lu, X.; He, C., Some recent developments of polyhedral oligomeric silsesquioxane (POSS)-based polymeric materials. *Journal of Materials Chemistry* **2011,** *21* (9), 2775-2782.

(137) Mya, K. Y.; Lin, E. M. J.; Gudipati, C. S.; Shen, L.; He, C., Time-Dependent Polymerization Kinetic Study and the Properties of Hybrid Polymers with Functional Silsesquioxanes. *The Journal of Physical Chemistry B* **2010,** *114* (28), 9119-9127.

(138) Pan, A.; Yang, S.; He, L.; Zhao, X., Star-shaped POSS diblock copolymers and their self-assembled films. *RSC Advances* **2014,** *4* (53), 27857-27866.

(139) Kango, S.; Kalia, S.; Celli, A.; Njuguna, J.; Habibi, Y.; Kumar, R., Surface modification of inorganic nanoparticles for development of organic–inorganic nanocomposites—A review. *Progress in Polymer Science* **2013,** *38* (8), 1232-1261.

(140) Que, Y.; Feng, C.; Zhang, S.; Huang, X., Stability and catalytic activity of PEG-b-PS-capped gold nanoparticles: a matter of PS chain length. *The Journal of Physical Chemistry C* **2015,** *119* (4), 1960-1970.

(141) He, J.; Liu, Y.; Babu, T.; Wei, Z.; Nie, Z., Self-assembly of inorganic nanoparticle vesicles and tubules driven by tethered linear block copolymers. *Journal of the American Chemical Society* **2012,** *134* (28), 11342-11345.

(142) Wang, J.; Li, W.; Zhu, J., Encapsulation of inorganic nanoparticles into block copolymer micellar aggregates: Strategies and precise localization of nanoparticles. *Polymer* **2014,** *55* (5), 1079-1096.

(143) Li, W.; Liu, S.; Deng, R.; Wang, J.; Nie, Z.; Zhu, J., A simple route to improve inorganic nanoparticles loading efficiency in block copolymer micelles. *Macromolecules* **2013,** *46* (6), 2282-2291.

(144) He, J.; Huang, X.; Li, Y.-C.; Liu, Y.; Babu, T.; Aronova, M. A.; Wang, S.; Lu, Z.; Chen, X.; Nie, Z., Self-Assembly of Amphiphilic Plasmonic Micelle-Like Nanoparticles in Selective Solvents. *Journal of the American Chemical Society* **2013,** *135* (21), 7974-7984.

(145) Liu, Y.; Li, Y.; He, J.; Duelge, K. J.; Lu, Z.; Nie, Z., Entropy-Driven Pattern Formation of Hybrid Vesicular Assemblies Made from Molecular and Nanoparticle Amphiphiles. *Journal of the American Chemical Society* **2014,** *136* (6), 2602-2610.

(146) Wang, L.; Liu, Y.; He, J.; Hourwitz, M. J.; Yang, Y.; Fourkas, J. T.; Han, X.; Nie, Z., Continuous Microfluidic Self-Assembly of Hybrid Janus-Like Vesicular Motors: Autonomous Propulsion and Controlled Release. *Small* **2015,** *11* (31), 3762-3767.

(147) Choueiri, R. M.; Galati, E.; Thérien-Aubin, H.; Klinkova, A.; Larin, E. M.; Querejeta-Fernández, A.; Han, L.; Xin, H. L.; Gang, O.; Zhulina, E. B.; Rubinstein, M.; Kumacheva, E., Surface patterning of nanoparticles with polymer patches. *Nature* **2016,** *538* (7623), 79-83.

(148) Kao, J.; Thorkelsson, K.; Bai, P.; Rancatore, B. J.; Xu, T., Toward functional nanocomposites: taking the best of nanoparticles, polymers, and small molecules. *Chemical Society Reviews* **2013,** *42* (7), 2654-2678.

(149) Yang, S.; Zhang, F.; Tai, J.; Li, Y.; Yang, Y.; Wang, H.; Zhang, J.; Xie, Z.; Xu, B.; Zhong, H.; Liu, K.; Yang, B., A detour strategy for colloidally stable block-copolymer grafted MAPbBr3 quantum dots in water with long photoluminescence lifetime. *Nanoscale* **2018,** *10* (13), 5820-5826.

(150) Raja, S. N.; Bekenstein, Y.; Koc, M. A.; Fischer, S.; Zhang, D.; Lin, L.; Ritchie, R. O.; Yang, P.; Alivisatos, A. P., Encapsulation of Perovskite Nanocrystals into Macroscale Polymer Matrices: Enhanced Stability and Polarization. *ACS Applied Materials & Interfaces* **2016,** *8* (51), 35523-35533.

(151) Chen, W.; Zhou, M.; Liu, Y.; Yu, X.; Pi, C.; Yang, Z.; Zhang, H.; Liu, Z.; Wang, T.; Qiu, J.; Yu, S. F.; Yang, Y.; Xu, X., All-Inorganic Perovskite Polymer–Ceramics for Flexible and Refreshable X-Ray Imaging. *Advanced Functional Materials* **2022,** *32* (2), 2107424.

(152) Hintermayr, V. A.; Lampe, C.; Löw, M.; Roemer, J.; Vanderlinden, W.; Gramlich, M.; Böhm, A. X.; Sattler, C.; Nickel, B.; Lohmüller, T.; Urban, A. S., Polymer Nanoreactors Shield Perovskite Nanocrystals from Degradation. *Nano Letters* **2019,** *19* (8), 4928-4933.

(153) Hui, L. S.; Beswick, C.; Getachew, A.; Heilbrunner, H.; Liang, K.; Hanta, G.; Arbi, R.; Munir, M.; Dawood, H.; Isik Goktas, N.; LaPierre, R.; Scharber, M. C.; Sariciftci, N. S.; Turak, A., Reverse Micelle Templating Route to Ordered Monodispersed Spherical Organo-Lead Halide Perovskite Nanoparticles for Light Emission. *ACS Applied Nano Materials* **2019,** *2* (7), 4121-4132.

(154) Zhang, K.; Fan, W.; Yao, T.; Wang, S.; Yang, Z.; Yao, J.; Xu, L.; Song, J., Polymer-Surface-Mediated Mechanochemical Reaction for Rapid and Scalable Manufacture of Perovskite QD Phosphors. *Advanced Materials* **2024,** *n/a* (n/a), 10.

(155) Děcká, K.; Pagano, F.; Frank, I.; Kratochwil, N.; Mihóková, E.; Auffray, E.; Čuba, V., Timing performance of lead halide perovskite nanoscintillators embedded in a polystyrene matrix. *Journal of Materials Chemistry C* **2022,** *10* (35), 12836-12843.

(156) Tomanová, K.; Suchá, A.; Mihóková, E.; Procházková, L.; Jakubec, I.; Turtos, R. M.; Gundacker, S.; Auffray, E.; Čuba, V., CsPbBr$_3$ Thin Films on LYSO:Ce Substrates. *IEEE Transactions on Nuclear Science* **2020,** *67* (6), 933-938.

(157) Děcká, K.; Král, J.; Hájek, F.; Průša, P.; Babin, V.; Mihóková, E.; Čuba, V. Scintillation Response Enhancement in Nanocrystalline Lead Halide Perovskite Thin Films on Scintillating Wafers *Nanomaterials* [Online], 2022, p. 14. https://www.mdpi.com/2079-4991/12/1/14.

(158) O'Neill, J.; Braddock, I.; Crean, C.; Ghosh, J.; Masteghin, M.; Richards, S.; Wilson, M.; Sellin, P., Development and characterisation of caesium lead halide perovskite nanocomposite scintillators for X-ray detection. *Frontiers in Physics* **2023,** *10*, 8.

(159) Braddock, I. H. B.; Al Sid Cheikh, M.; Ghosh, J.; Mulholland, R. E.; O'Neill, J. G.; Stolojan, V.; Crean, C.; Sweeney, S. J.; Sellin, P. J., Formamidinium Lead Halide Perovskite Nanocomposite Scintillators. *Nanomaterials* **2022,** *12* (13), 2141.

(160) Liu, X.-Y.; Pilania, G.; Talapatra, A. A.; Stanek, C. R.; Uberuaga, B. P., Band-Edge Engineering To Eliminate Radiation-Induced Defect States in Perovskite Scintillators. *ACS Applied Materials & Interfaces* **2020,** *12* (41), 46296-46305.

(161) Erman, B.; Flory, P. J., Critical phenomena and transitions in swollen polymer networks and in linear macromolecules. *Macromolecules* **1986,** *19* (9), 2342-2353.

(162) Freitas, S.; Merkle, H. P.; Gander, B., Microencapsulation by solvent extraction/evaporation: reviewing the state of the art of microsphere preparation process technology. *Journal of controlled release* **2005,** *102* (2), 313-332.

(163) Li, M.; Rouaud, O.; Poncelet, D., Microencapsulation by solvent evaporation: State of the art for process engineering approaches. *International Journal of pharmaceutics* **2008,** *363* (1-2), 26-39.

(164) Wang, Y.; He, J.; Chen, H.; Chen, J.; Zhu, R.; Ma, P.; Towers, A.; Lin, Y.; Gesquiere, A. J.; Wu, S.-T.; Dong, Y., Ultrastable, Highly Luminescent Organic–Inorganic Perovskite–Polymer Composite Films. *Advanced Materials* **2016,** *28* (48), 10710-10717.

(165) Zhao, H.; Benetti, D.; Tong, X.; Zhang, H.; Zhou, Y.; Liu, G.; Ma, D.; Sun, S.; Wang, Z. M.; Wang, Y.; Rosei, F., Efficient and stable tandem luminescent solar concentrators based on carbon dots and perovskite quantum dots. *Nano Energy* **2018,** *50*, 756-765.

(166) Zhao, H.; Zhou, Y.; Benetti, D.; Ma, D.; Rosei, F., Perovskite quantum dots integrated in large-area luminescent solar concentrators. *Nano Energy* **2017,** *37*, 214-223.

(167) Meinardi, F.; Ehrenberg, S.; Dhamo, L.; Carulli, F.; Mauri, M.; Bruni, F.; Simonutti, R.; Kortshagen, U.; Brovelli, S., Highly efficient luminescent solar concentrators based on earth-abundant indirect-bandgap silicon quantum dots. *Nature Photonics* **2017,** *11* (3), 177-185.

(168) Moraitis, P.; Schropp, R. E. I.; van Sark, W. G. J. H. M., Nanoparticles for Luminescent Solar Concentrators - A review. *Optical Materials* **2018,** *84*, 636-645.

(169) Dhamo, L.; Carulli, F.; Nickl, P.; Wegner, K. D.; Hodoroaba, V.-D.; Würth, C.; Brovelli, S.; Resch-Genger, U., Efficient Luminescent Solar Concentrators Based on Environmentally Friendly Cd-Free Ternary AIS/ZnS Quantum Dots. *Advanced Optical Materials* **2021,** *9* (17), 2100587.

(170) Mattiello, S.; Sanzone, A.; Bruni, F.; Gandini, M.; Pinchetti, V.; Monguzzi, A.; Facchinetti, I.; Ruffo, R.; Meinardi, F.; Mattioli, G.; Sassi, M.; Brovelli, S.; Beverina, L., Chemically Sustainable Large Stokes Shift Derivatives for High-Performance Large-Area Transparent Luminescent Solar Concentrators. *Joule* **2020,** *4* (9), 1988-2003.

(171) Rafiee, M.; Chandra, S.; Ahmed, H.; McCormack, S. J., An overview of various configurations of Luminescent Solar Concentrators for photovoltaic applications. *Optical Materials* **2019,** *91*, 212-227.

(172) Wei, M.; de Arquer, F. P. G.; Walters, G.; Yang, Z.; Quan, L. N.; Kim, Y.; Sabatini, R.; Quintero-Bermudez, R.; Gao, L.; Fan, J. Z.; Fan, F.; Gold-Parker, A.; Toney, M. F.; Sargent, E. H., Ultrafast narrowband exciton routing within layered perovskite nanoplatelets enables low-loss luminescent solar concentrators. *Nature Energy* **2019,** *4* (3), 197-205.

(173) Xin, Y.; Zhao, H.; Zhang, J., Highly Stable and Luminescent Perovskite–Polymer Composites from a Convenient and Universal Strategy. *ACS Applied Materials & Interfaces* **2018,** *10* (5), 4971-4980.

(174) Rodová, M.; Brožek, J.; Knížek, K.; Nitsch, K., Phase transitions in ternary caesium lead bromide. *Journal of Thermal Analysis and Calorimetry* **2003,** *71* (2), 667-673.

(175) Zhang, Q.; Li, Z.; Liu, M.; Kong, L.; Zheng, W.; Wang, B.; Li, L., Bifunctional Passivation Strategy to Achieve Stable $CsPbBr_3$ Nanocrystals with Drastically Reduced Thermal-Quenching. *The Journal of Physical Chemistry Letters* **2020,** *11* (3), 993-999.

(176) Hu, X.; Wu, Y.; Wang, Y.; Xu, L.; Zhang, S.; Wang, J.; Wu, K.; Liu, Y.; Li, Y.; Li, X., Surface Anchoring-Induced Robust Luminescence Thermal Quenching Suppression in Shell-Free Perovskite Nanocrystals. *Advanced Optical MAterials* **2022,** *10* (22), 9.

(177) Zhang, Q.; He, M.; Wan, Q.; Zheng, W.; Liu, M.; Zhang, C.; Liao, X.; Zhan, W.; Kong, L.; Guo, X.; Li, L., Suppressing thermal quenching of lead halide perovskite nanocrystals by constructing a wide-bandgap surface layer for achieving thermally stable white light-emitting diodes. *Chemical Science* **2022,** *13* (13), 3719-3727.

(178) Xu, J.; Atme, A.; Marques Martins, A. F.; Jung, K.; Boyer, C., Photoredox catalyst-mediated atom transfer radical addition for polymer functionalization under visible light. *Polymer Chemistry* **2014,** *5* (10), 3321-3325.

(179) Bellotti, V.; Simonutti, R., New Light in Polymer Science: Photoinduced Reversible Addition-Fragmentation Chain Transfer Polymerization (PET-RAFT) as Innovative Strategy for the Synthesis of Advanced Materials. *Polymers* **2021,** *13* (7), 1119.

(180) Jin, X.; Ma, K.; Chakkamalayath, J.; Morsby, J.; Gao, H., In Situ Photocatalyzed Polymerization to Stabilize Perovskite Nanocrystals in Protic Solvents. *ACS Energy Letters* **2022,** *7* (2), 610-616.

(181) Zhu, Y.; Liu, Y.; Miller, K. A.; Zhu, H.; Egap, E., Lead Halide Perovskite Nanocrystals as Photocatalysts for PET-RAFT Polymerization under Visible and Near-Infrared Irradiation. *ACS Macro Letters* **2020,** *9* (5), 725-730.

(182) Allegrezza, M. L.; Konkolewicz, D., PET-RAFT Polymerization: Mechanistic Perspectives for Future Materials. *ACS Macro Letters* **2021,** *10* (4), 433-446.

(183) Moretti, F.; Patton, G.; Belsky, A.; Petrosyan, A. G.; Dujardin, C., Deep traps can reduce memory effects of shallower ones in scintillators. *Physical Chemistry Chemical Physics* **2016,** *18* (2), 1178-1184.

(184) Galunov, N.; Gryn, D.; Karavaeva, N.; Khromiuk, I.; Lazarev, I.; Navozenko, O.; Naumenko, A.; Tarasenko, O.; Yashchuk, V., Delayed radioluminescence of some heterostructured organic scintillators. *Journal of Luminescence* **2020,** *226*, 117477.

(185) Mandal, S.; Mukherjee, S.; De, C. K.; Roy, D.; Ghosh, S.; Mandal, P. K., Extent of Shallow/Deep Trap States beyond the Conduction Band Minimum in Defect-Tolerant $CsPbBr_3$ Perovskite Quantum Dot: Control over the Degree of Charge Carrier Recombination. *The Journal of Physical Chemistry Letters* **2020,** *11* (5), 1702-1707.

(186) Li, Z.; Hu, Q.; Tan, Z.; Yang, Y.; Leng, M.; Liu, X.; Ge, C.; Niu, G.; Tang, J., Aqueous Synthesis of Lead Halide Perovskite Nanocrystals with High Water Stability and Bright Photoluminescence. *ACS Applied Materials & Interfaces* **2018,** *10* (50), 43915-43922.

(187) Cho, S.; Kim, S.; Kim, J.; Jo, Y.; Ryu, I.; Hong, S.; Lee, J.-J.; Cha, S.; Nam, E. B.; Lee, S. U.; Noh, S. K.; Kim, H.; Kwak, J.; Im, H., Hybridisation of perovskite nanocrystals with organic molecules for highly efficient liquid scintillators. *Light: Science & Applications* **2020,** *9* (1), 156.

(188) Yang, H.; Li, H.; Yuan, R.; Chen, J.; Zhao, J.; Wang, S.; Liu, Y.; Li, Q.; Zhang, Z., A novel scintillation screen for achieving high-energy ray detection with fast and full-color emission. *Journal of Materials Chemistry C* **2021,** *9* (25), 7905-7909.

(189) Wei, J.-H.; Wang, X.-D.; Liao, J.-F.; Kuang, D.-B., High Photoluminescence Quantum Yield (>95%) of $MAPbBr_3$ Nanocrystals via Reprecipitation from Methylamine-MAPbBr3 Liquid. *ACS Applied Electronic Materials* **2020,** *2* (9), 2707-2715.

(190) Pauwels, K.; Douissard, P.-A., Indirect X-ray detectors with single-photon sensitivity. *Journal of Synchrotron Radiation* **2022,** *29* (6), 1394-1406.

(191) Naresh, V.; Singh, S.; Soh, H.; Lee, J.; Lee, N., Dual-phase CsPbBr$_3$–CsPb$_2$Br$_5$ perovskite scintillator for sensitive X-ray detection and imaging. *Materials Today Nano* **2023,** *23*, 100364.

(192) Lian, H.; Zhang, W.; Zou, R.; Gu, S.; Kuang, R.; Zhu, Y.; Zhang, X.; Ma, C.-G.; Wang, J.; Li, Y., Aqueous-Based Inorganic Colloidal Halide Perovskites Customizing Liquid Scintillators. *Advanced Materials* **2023,** *35* (51), 2304743.

(193) Liu, M.; Huang, L.; Yuan, D.; Li, Z.; Teng, Y.; Zhang, J.; Huang, S.; Liu, B., Perovskite Nanocrystals and Dyes for High-Efficiency Liquid Scintillator Counters To Detect Radiation. *ACS Applied Nano Materials* **2023,** *6* (1), 370-378.

(194) Dierks, H.; Zhang, Z.; Lamers, N.; Wallentin, J., 3D X-ray microscopy with a CsPbBr$_3$ nanowire scintillator. *Nano Research* **2022,** *16* (1), 1084-1089.

(195) Birowosuto, M. D.; Maddalena, F.; Xie, A.; Witkowski, M. E.; Makowski, M.; Drozdowski, W.; Coquet, P.; Christophe, D.; Cuong, D. In *Scintillators from solution-processable perovskite halide single crystals or quantum dots: the good, the bad, and the ugly*, Proc.SPIE, 2020; p 1149415.

(196) Yu, H.; Chen, T.; Han, Z.; Fan, J.; Pei, Q., Liquid Scintillators Loaded with up to 40 Weight Percent Cesium Lead Bromide Quantum Dots for Gamma Scintillation. *ACS Applied Nano Materials* **2022,** *5* (10), 14572-14581.

(197) Lü, Z.-W.; Wei, G.-X.; Wang, H.-Q.; Guan, Y.; Jiang, N.; Liu, Y.-Y.; Li, Z.; Qin, H.; Liu, H.-Q., New flexible CsPbBr$_3$-based scintillator for X-ray tomography. *Nuclear Science and Techniques* **2022,** *33* (8), 98.

(198) Zaffalon, M. L.; Wu, Y.; Cova, F.; Gironi, L.; Li, X.; Pinchetti, V.; Liu, Y.; Imran, M.; Cemmi, A.; Di Sarcina, I.; Manna, L.; Zeng, H.; Brovelli, S., Zero-Dimensional Gua$_3$SbCl$_6$ Crystals as Intrinsically Reabsorption-Free Scintillators for Radiation Detection. *Advanced Functional Materials* **2023,** *n/a* (n/a), 2305564.

(199) Amsler, C.; Grögler, D.; Joffrain, W.; Lindelöf, D.; Marchesotti, M.; Niederberger, P.; Pruys, H.; Regenfus, C.; Riedler, P.; Rotondi, A., Temperature dependence of pure CsI: scintillation light yield and decay time. *Nuclear Instruments and Methods in Physics Research Section A: Accelerators, Spectrometers, Detectors and Associated Equipment* **2002,** *480* (2), 494-500.

(200) Visser, R.; Dorenbos, P.; Eijk, C. W. E. v.; Hollander, R. W.; Schotanus, P., Scintillation properties of Ce$^{3+}$ doped BaF$_2$ crystals. *IEEE Transactions on Nuclear Science* **1991,** *38* (2), 178-183.

(201) Wang, Z.; Dujardin, C.; Freeman, M. S.; Gehring, A. E.; Hunter, J. F.; Lecoq, P.; Liu, W.; Melcher, C. L.; Morris, C. L.; Nikl, M.; Pilania, G.; Pokharel, R.; Robertson, D. G.; Rutstrom, D. J.; Sjue, S. K.; Tremsin, A. S.; Watson, S. A.; Wiggins, B. W.; Winch, N. M.; Zhuravleva, M., Needs, Trends, and Advances in Scintillators for Radiographic Imaging and Tomography. *IEEE Transactions on Nuclear Science* **2023,** *70* (7), 1244-1280.

(202) Hao, S.; Liu, X.; Li, Q.; Gu, M.; Cheng, S.; Zhao, J., Ultrafast Decay, Ultrahigh Spatial Resolution, and Stable γ-CuI Single Crystal Treated by Iodine Annealing and SiO$_2$ Coating. *ACS Applied Materials & Interfaces* **2023,** *15* (37), 44493-44502.

(203) Yue, S.; Gu, M.; Liu, X.; Li, F.; Liu, S.; Zhang, X.; Zhang, J.; Liu, B.; Huang, S.; Ni, C., Optimization of crystal growth and properties of γ-CuI ultrafast scintillator by the addition of LiI. *Materials Research Bulletin* **2018,** *106*, 228-233.

(204) Nassalski, A.; Kapusta, M.; Batsch, T.; Wolski, D.; Mockel, D.; Enghardt, W.; Moszynski, M. In *Comparative study of scintillators for PET/CT detectors*, IEEE Nuclear Science Symposium Conference Record, 2005, 23-29 Oct. 2005; 2005; pp 2823-2829.

(205) Mouhti, I.; Elanique, A.; Messous, M. Y.; Benahmed, A.; McFee, J. E.; Elgoub, Y.; Griffith, P., Characterization of CsI(Tl) and LYSO(Ce) scintillator detectors by measurements and Monte Carlo simulations. *Applied Radiation and Isotopes* **2019,** *154*, 108878.

(206) Nikl, M.; Yoshikawa, A.; Kamada, K.; Nejezchleb, K.; Stanek, C. R.; Mares, J. A.; Blazek, K., Development of LuAG-based scintillator crystals – A review. *Progress in Crystal Growth and Characterization of Materials* **2013,** *59* (2), 47-72.

(207) Mares, J. A.; Nikl, M.; Beitlerova, A.; Horodysky, P.; Blazek, K.; Bartos, K.; Ambrosio, C. D., Scintillation Properties of $Ce^{3+}$- and $Pr^{3+}$-Doped LuAG, YAG and Mixed $Lu_xY_{1-x}AG$ Garnet Crystals. *IEEE Transactions on Nuclear Science* **2012,** *59* (5), 2120-2125.

(208) Zhou, Y.; Chen, J.; Bakr, O. M.; Mohammed, O. F., Metal Halide Perovskites for X-ray Imaging Scintillators and Detectors. *ACS Energy Letters* **2021,** *6* (2), 739-768.

(209) Rodà, C.; Fasoli, M.; Zaffalon, M. L.; Cova, F.; Pinchetti, V.; Shamsi, J.; Abdelhady, A. L.; Imran, M.; Meinardi, F.; Manna, L.; Vedda, A.; Brovelli, S., Understanding Thermal and A-Thermal Trapping Processes in Lead Halide Perovskites Towards Effective Radiation Detection Schemes. *Advanced Functional Materials* **2021,** *31* (43), 2104879.

(210) Cova, F.; Erroi, A.; Zaffalon, M. L.; Cemmi, A.; Di Sarcina, I.; Perego, J.; Monguzzi, A.; Comotti, A.; Rossi, F.; Carulli, F.; Brovelli, S., Scintillation Properties of $CsPbBr_3$ Nanocrystals Prepared by Ligand-Assisted Reprecipitation and Dual Effect of Polyacrylate Encapsulation toward Scalable Ultrafast Radiation Detectors. *Nano Letters* **2024**.

(211) Xie, A.; Maddalena, F.; Witkowski, M. E.; Makowski, M.; Mahler, B.; Drozdowski, W.; Springham, S. V.; Coquet, P.; Dujardin, C.; Birowosuto, M. D.; Dang, C., Library of Two-Dimensional Hybrid Lead Halide Perovskite Scintillator Crystals. *Chemistry of Materials* **2020,** *32* (19), 8530-8539.

(212) Kang, J.; Wang, L.-W., High Defect Tolerance in Lead Halide Perovskite $CsPbBr_3$. *The Journal of Physical Chemistry Letters* **2017,** *8* (2), 489-493.

(213) Bodnarchuk, M. I.; Boehme, S. C.; ten Brinck, S.; Bernasconi, C.; Shynkarenko, Y.; Krieg, F.; Widmer, R.; Aeschlimann, B.; Günther, D.; Kovalenko, M. V.; Infante, I., Rationalizing and Controlling the Surface Structure and Electronic Passivation of Cesium Lead Halide Nanocrystals. *ACS Energy Letters* **2019,** *4* (1), 63-74.

(214) Kumar, V.; Luo, Z. A Review on X-ray Excited Emission Decay Dynamics in Inorganic Scintillator Materials *Photonics* [Online], 2021, p. 71. https://www.mdpi.com/2304-6732/8/3/71.

(215) Lang, F.; Nickel, N. H.; Bundesmann, J.; Seidel, S.; Denker, A.; Albrecht, S.; Brus, V. V.; Rappich, J.; Rech, B.; Landi, G.; Neitzert, H. C., Radiation Hardness and Self-Healing of Perovskite Solar Cells. *Advanced Materials* **2016,** *28* (39), 8726-8731.

(216) Lang, F.; Shargaieva, O.; Brus, V. V.; Neitzert, H. C.; Rappich, J.; Nickel, N. H., Influence of Radiation on the Properties and the Stability of Hybrid Perovskites. *Advanced Materials* **2018,** *30* (3), 1702905.

(217) Kirmani, A. R.; Durant, B. K.; Grandidier, J.; Haegel, N. M.; Kelzenberg, M. D.; Lao, Y. M.; McGehee, M. D.; McMillon-Brown, L.; Ostrowski, D. P.; Peshek, T. J.; Rout, B.; Sellers, I. R.;

Steger, M.; Walker, D.; Wilt, D. M.; VanSant, K. T.; Luther, J. M., Countdown to perovskite space launch: Guidelines to performing relevant radiation-hardness experiments. *Joule* **2022,** *6* (5), 1015-1031.

(218) Miyazawa, Y.; Ikegami, M.; Chen, H.-W.; Ohshima, T.; Imaizumi, M.; Hirose, K.; Miyasaka, T., Tolerance of Perovskite Solar Cell to High-Energy Particle Irradiations in Space Environment. *iScience* **2018,** *2,* 148-155.

(219) Brus, V. V.; Lang, F.; Bundesmann, J.; Seidel, S.; Denker, A.; Rech, B.; Landi, G.; Neitzert, H. C.; Rappich, J.; Nickel, N. H., Defect Dynamics in Proton Irradiated $CH_3NH_3PbI_3$ Perovskite Solar Cells. *Advanced Electronic Materials* **2017,** *3* (2), 1600438.

(220) Wu, X.; Guo, Z.; Zhu, S.; Zhang, B.; Guo, S.; Dong, X.; Mei, L.; Liu, R.; Su, C.; Gu, Z., Ultrathin, Transparent, and High Density Perovskite Scintillator Film for High Resolution X-Ray Microscopic Imaging. *Advanced Science* **2022,** *9* (17), 2200831.

(221) Yang, L.; Zhang, H.; Zhou, M.; Zhao, L.; Chen, W.; Wang, T.; Yu, X.; Zhou, D.; Qiu, J.; Xu, X., High-Stable X-ray Imaging from All-Inorganic Perovskite Nanocrystals under a High Dose Radiation. *The Journal of Physical Chemistry Letters* **2020,** *11* (21), 9203-9209.

(222) Todesco, E.; Bajas, H.; Bajko, M.; Ballarino, A.; Bermudez, S. I.; Bordini, B.; Bottura, L.; De Rijk, G.; Devred, A.; Duarte Ramos, D.; Duda, M.; Ferracin, P.; Fessia, P.; Fleiter, J.; Fiscarelli, L.; Foussat, A.; Kirby, G.; Mangiarotti, F.; Mentink, M.; Milanese, A.; Musso, A.; Parma, V.; Perez, J. C.; Prin, H.; Rossi, L.; Russenschuck, S.; Willering, G.; Enomoto, S.; Nakamoto, T.; Kimura, N.; Ogitsu, T.; Sugano, M.; Suzuki, K.; Wei, S.; Gong, L.; Wang, J.; Peng, Q.; Xu, Q.; Bersani, A.; Caiffi, B.; Fabbricatore, P.; Farinon, S.; Pampaloni, A.; Mariotto, S.; Prioli, M.; Sorbi, M.; Statera, M.; Garcia Matos, J.; Toral, F.; Ambrosio, G.; Apollinari, G.; Baldini, M.; Carcagno, R.; Feher, S.; Stoynev, S.; Chlachidze, G.; Marinozzi, V.; Lombardo, V.; Nobrega, F.; Strauss, T.; Yu, M.; Anerella, M.; Amm, K.; Joshi, P.; Muratore, J.; Schmalzle, J.; Wanderer, P.; Chen, D.; Gourlay, S.; Pong, I.; Prestemon, S.; Sabbi, G. L.; Cooley, L.; Felice, H., The High Luminosity LHC interaction region magnets towards series production. *Superconductor Science and Technology* **2021,** *34* (5), 053001.

(223) Olszacki, M.; Matusiak, M.; Augustyniak, I.; Knapkiewicz, P.; Dziuban, J.; Pons, P.; Debourg, E., Measurement of the High Gamma Radiation Dose Using The MEMS Based Dosimeter and Radiolisys Effect. *MicroMechanics Europe Workshop* **2013,** 31-35. .

(224) Zhou, Y.; Deng, Z.; Wang, B.; Li, P.; Li, L.; Han, W.; Huang, J.; Jia, W.; Ouyang, X.; Xu, Q.; Ostrikov, K., Nanocomposite scintillation perovskite-delignified wood photonic guides for X-ray imaging. *Chemical Engineering Journal* **2023,** *471,* 144431.

(225) Li, P.; Cheng, W.; Zhou, Y.; Zhao, D.; Liu, J.; Li, L.; Ouyang, X.; Liu, B.; Jia, W.; Xu, Q.; Ostrikov, K., Large Scale BN-perovskite Nanocomposite Aerogel Scintillator for Thermal Neutron Detection. *Advanced Materials* **2023,** *35* (25), 2209452.

(226) Chen, H.; Wang, Q.; Peng, G.; Wang, S.; Lei, Y.; Wang, H.; Yang, Z.; Sun, J.; Li, N.; Zhao, L.; Lan, W.; Jin, Z., Cesium Lead Halide Nanocrystals based Flexible X-Ray Imaging Screen and Visible Dose Rate Indication on Paper Substrate. *Advanced Optical Materials* **2022,** *10* (8), 2102790.

(227) Milotti, V.; Cacovich, S.; Ceratti, D. R.; Ory, D.; Barichello, J.; Matteocci, F.; Di Carlo, A.; Sheverdyaeva, P. M.; Schulz, P.; Moras, P., Degradation and Self-Healing of $FAPbBr_3$ Perovskite under Soft-X-Ray Irradiation. *Small Methods* **2023,** *7* (9), 2300222.

(228) Gao, L.; Tao, K.; Sun, J.-L.; Yan, Q., Gamma-Ray Radiation Stability of Mixed-Cation Lead Mixed-Halide Perovskite Single Crystals. *Advanced Optical Materials* **2022,** *10* (3), 2102069.

(229) Dujardin, C.; Hamel, M., Introduction—Overview on Plastic and Inorganic Scintillators. In *Plastic Scintillators: Chemistry and Applications*, Hamel, M., Ed. Springer International Publishing: Cham, 2021; pp 3-33.

(230) Conti, M., Focus on time-of-flight PET: the benefits of improved time resolution. *European Journal of Nuclear Medicine and Molecular Imaging* **2011,** *38* (6), 1147-57.

(231) Skrypnyk, T.; Viahin, O.; Bespalova, I.; Zelenskaya, O.; Tarasov, V.; Alekseev, V.; Yefimova, S.; Sorokin, O., Scintillation properties of composite films based on $CsPbBr_3$ nanocrystals embedded in PMMA. *Radiation Measurements* **2023,** *169*, 107028.

(232) Lecoq, P.; Morel, C.; Prior, J. O.; Visvikis, D.; Gundacker, S.; Auffray, E.; Križan, P.; Turtos, R. M.; Thers, D.; Charbon, E.; Varela, J.; de La Taille, C.; Rivetti, A.; Breton, D.; Pratte, J.-F.; Nuyts, J.; Surti, S.; Vandenberghe, S.; Marsden, P.; Parodi, K.; Benlloch, J. M.; Benoit, M., Roadmap toward the 10 ps time-of-flight PET challenge. *Physics in Medicine & Biology* **2020,** *65* (21), 21RM01.

(233) Auffray, E.; Frisch, B.; Geraci, F.; Ghezzi, A.; Gundacker, S.; Hillemanns, H.; Jarron, P.; Meyer, T.; Paganoni, M.; Pauwels, K.; Pizzichemi, M.; Lecoq, P., A Comprehensive & Systematic Study of Coincidence Time Resolution and Light Yield Using Scintillators of Different Size and Wrapping. *IEEE Transactions on Nuclear Science* **2013,** *60* (5), 3163-3171.

(234) Dujardin, C.; Auffray, E.; Bourret-Courchesne, E.; Dorenbos, P.; Lecoq, P.; Nikl, M.; Vasil'ev, A. N.; Yoshikawa, A.; Zhu, R.-Y., Needs, Trends, and Advances in Inorganic Scintillators. *IEEE Transactions on Nuclear Science* **2018,** *65* (8), 1977-1997.

(235) Saleem, T.; Ahmad, S.; Cizel, J.-B.; De La Taille, C.; Morenas, M.; Nadig, V.; Perez, F.; Schulz, V.; Gundacker, S.; Fleury, J., Study experimental time resolution limits of recent ASICs at Weeroc with different SiPMs and scintillators. *Journal of Instrumentation* **2023,** *18* (10), P10005.

(236) Joyce van, S.; Johan de, J.; Jenny, S.; Walter, N.; Paul van, S.; Rudi, D.; Ronald, B.; Antoon, W.; Ronald, B., Performance Characteristics of the Digital Biograph Vision PET/CT System. *Journal of Nuclear Medicine* **2019,** *60* (7), 1031.

(237) Lecoq, P., Pushing the Limits in Time-of-Flight PET Imaging. *IEEE Transactions on Radiation and Plasma Medical Sciences* **2017,** *1* (6), 473-485.

(238) Shao, Y., A new timing model for calculating the intrinsic timing resolution of a scintillator detector. *Physics in Medicine & Biology* **2007,** *52* (4), 1103.

(239) Lecoq, P.; Auffray, E.; Brunner, S. E.; Hillemanns, H.; Jarron, P.; Knapitsch, A.; Meyer, T.; Powolny, F., Factors Influencing Time Resolution of Scintillators and Ways to Improve Them. *IEEE Transactions on Nuclear Science* **2010,** *57* (5), 2411-2416.

(240) Kratochwil, N.; Gundacker, S.; Auffray, E., A roadmap for sole Cherenkov radiators with SiPMs in TOF-PET. *Physics in Medicine & Biology* **2021,** *66* (19), 195001.

(241) Omelkov, S. I.; Nagirnyi, V.; Gundacker, S.; Spassky, D. A.; Auffray, E.; Lecoq, P.; Kirm, M., Scintillation yield of hot intraband luminescence. *Journal of Luminescence* **2018,** *198*, 260-271.

(242) Gundacker, S.; Pots, R. H.; Nepomnyashchikh, A.; Radzhabov, E.; Shendrik, R.; Omelkov, S.; Kirm, M.; Acerbi, F.; Capasso, M.; Paternoster, G.; Mazzi, A.; Gola, A.; Chen, J.; Auffray, E., Vacuum ultraviolet silicon photomultipliers applied to $BaF_2$ cross-luminescence detection for high-rate ultrafast timing applications. *Physics in Medicine & Biology* **2021,** *66* (11), 114002.

(243) Tao, L.; Coffee, R. N.; Jeong, D.; Levin, C. S., Ionizing photon interactions modulate the optical properties of crystals with femtosecond scale temporal resolution. *Physics in Medicine & Biology* **2021,** *66* (4), 045032.

(244) Tomanová, K.; Čuba, V.; Brik, M. G.; Mihóková, E.; Martinez Turtos, R.; Lecoq, P.; Auffray, E.; Nikl, M., On the structure, synthesis, and characterization of ultrafast blue-emitting $CsPbBr_3$ nanoplatelets. *APL Materials* **2019,** *7* (1), 011104.

(245) Turtos, R. M.; Gundacker, S.; Omelkov, S.; Auffray, E.; Lecoq, P., Light yield of scintillating nanocrystals under X-ray and electron excitation. *Journal of Luminescence* **2019,** *215*, 116613.

(246) Konstantinou, G.; Lecoq, P.; Benlloch, J. M.; Gonzalez, A. J., Metascintillators for Ultrafast Gamma Detectors: A Review of Current State and Future Perspectives. *IEEE Transactions on Radiation and Plasma Medical Sciences* **2022,** *6* (1), 5-15.

(247) Pagano, F.; Král, J.; Decká, K.; Pizzichemi, M.; Mihóková, E.; Cuba, V.; Auffray, E., Nanocrystalline Lead Halide Perovskites to Boost Time-of-Flight Performance of Medical Imaging Detectors. *Advanced Materials Interfaces* **2024,** *n/a* (n/a), 2300659.

(248) Pagano, F.; Kratochwil, N.; Salomoni, M.; Pizzichemi, M.; Paganoni, M.; Auffray, E., Advances in heterostructured scintillators: toward a new generation of detectors for TOF-PET. *Physics in Medicine & Biology* **2022,** *67* (13), 135010.